\documentclass[english,10pt]{amsart}
\UseRawInputEncoding
\usepackage{amsfonts}
\usepackage{amssymb, amsmath}
\usepackage{amsthm}
\usepackage{mathtools}
\usepackage{dsfont}
\usepackage{bm}
\usepackage{verbatim}
\usepackage{xcolor}
\usepackage{url}
\usepackage{subcaption}
\usepackage{float}
\usepackage{caption}
\usepackage[pdftex]{hyperref}
\hypersetup{plainpages=false,unicode=true,pdffitwindow=true}
\usepackage{graphicx}
\usepackage{comment}
\usepackage{hyperref}
\usepackage{shuffle}
\usepackage{algorithm}
\usepackage{algpseudocode}
\usepackage{multirow}

\theoremstyle{plain}

\theoremstyle{definition}

\newcommand{\R}{\mathbb{R}}

\newcommand{\E}{\mathbb{E}}

\newcommand{\bx}{\bm{x}}
\newcommand{\by}{\bm{y}}
\newcommand{\bz}{\bm{z}}

\newcommand{\bw}{\bm{w}}
\newcommand{\bv}{\bm{v}}

\newcommand{\abs}[1]{\left\lvert #1 \right\rvert}
\newcommand{\id}{\textup{Id}}
\newcommand{\diag}{\textup{diag}}

\newcommand{\tol}{\textup{TOL}}

\newcommand{\dd}{\mathrm{d}}
\newcommand{\sdd}{\,\dd}

\newcommand{\vertiii}[1]{{\left\vert\kern-0.25ex\left\vert\kern-0.25ex\left\vert #1 
    \right\vert\kern-0.25ex\right\vert\kern-0.25ex\right\vert}}

\usepackage{todonotes}

\numberwithin{equation}{section}

\usepackage[backend=bibtex,style=alphabetic]{biblatex}

\bibliography{literature.bib}

\title[Simulation of rough Heston]{Efficient Option Pricing in the Rough Heston Model Using Weak Simulation Schemes}

\author{Christian Bayer}
\address{Weierstrass Institute, Mohrenstraße 39, 10117 Berlin, Germany}
\email{christian.bayer@wias-berlin.de}

\author{Simon Breneis}
\address{Weierstrass Institute, Mohrenstraße 39, 10117 Berlin, Germany}
\email{simon.breneis@wias-berlin.de}

\makeatletter
\@namedef{subjclassname@2020}{\textup{2020} Mathematics Subject Classification}
\makeatother

\date{\today}
\subjclass[2020]{91G60, 91G20}
\keywords{Rough Heston model, Markovian approximations, Simulation, Weak error, Bermudan options}

\thanks{C.B. and S.B. gratefully acknowledge the support of the DFG through the IRTG 2544. The authors would also like to thank Eduardo Abi Jaber for encouraging them to consider Hurst parameters $H<0$.}

\begin{document}
\maketitle

\begin{abstract}
     We provide an efficient and accurate simulation scheme for the rough Heston model in the standard ($H>0$) as well as the hyper-rough regime ($H > -1/2$). The scheme is based on low-dimensional Markovian approximations of the rough Heston process derived in [Bayer and Breneis, arXiv:2309.07023], and provides weak approximation to the rough Heston process. Numerical experiments show that the new scheme exhibits second order weak convergence, while the computational cost increases linear with respect to the number of time steps. In comparison, existing schemes based on discretization of the underlying stochastic Volterra integrals such as Gatheral's HQE scheme show a quadratic dependence of the computational cost. Extensive numerical tests for standard and path-dependent European options and Bermudan options show the method's accuracy and efficiency.

\end{abstract}

\section{Introduction}

The Heston model, first considered in \cite{heston1993closed}, is a popular option pricing model in mathematical finance. Its dynamics are given by 
\begin{align}
\dd \widetilde S_t &= \sqrt{\widetilde V_t} \widetilde S_t \left(\rho \sdd W_t + \sqrt{1-\rho^2} \sdd B_t\right),\qquad \widetilde S_0 = S_0,\label{eqn:HestonStock}\\
\dd \widetilde V_t &= (\theta - \lambda \widetilde V_t) \sdd t + \nu\sqrt{\widetilde V_t} \sdd W_t,\qquad \widetilde V_0 = V_0,\label{eqn:HestonVol}
\end{align}
where $\theta, \lambda, \nu > 0$, $\rho\in[-1, 1]$, and where $(B_t,W_t)$ is a two-dimensional standard Brownian motion.

One reason for its popularity is that the process $(\widetilde S, \widetilde V)$ is affine, and its characteristic function can thus be given explicitly, see again \cite{heston1993closed}. This enables very fast pricing of e.g. European options by Fourier inversion. However, one of its drawbacks is that there are several stylized facts of the market that the Heston model does not capture, for example the explosion of the implied volatility skew, see also \cite{el2019characteristic}. To resolve these problems, the rough Heston model was introduced in \cite{el2019characteristic, euch2018perfect}. Its dynamics are given by
\begin{align}
\dd S_t &= \sqrt{V_t} S_t \left(\rho \sdd W_t + \sqrt{1-\rho^2} \sdd B_t\right),\qquad S_0 = S_0,\label{eqn:RHestonStock}\\
V_t &= V_0 + \int_0^t K(t-s) (\theta - \lambda V_s) \sdd s + \int_0^t K(t-s) \nu\sqrt{V_s} \sdd W_s,\label{eqn:RHestonVol}
\end{align}
where $K$ is the fractional kernel $$K(t) \coloneqq \frac{t^{H-1/2}}{\Gamma(H+1/2)} = \int_0^\infty e^{-xt} \mu(\dd x),\quad \mu(\dd x) \coloneqq \frac{x^{-H-1/2} \sdd x}{\Gamma(H+1/2)\Gamma(1/2-H)},$$ for all $t > 0$, with Hurst parameter $H\in(-1/2, 1/2)$, and where $\Gamma$ is the Gamma-function. The use of rough volatility models like the rough Heston model \eqref{eqn:RHestonStock}-\eqref{eqn:RHestonVol} or the rough Bergomi model \cite{bayer2016pricing} is by now an established paradigm for modelling equity markets, as it provides excellent fits to market data, see e.g. \cite{roughvolbook}.

While the rough Heston model resolves several of the shortcomings of the (standard) Heston model, the singular Volterra-type dynamics of the volatility process $V$ in \eqref{eqn:RHestonVol} introduce significant challenges both in the theoretical analysis of this model, as well as in simulation and option pricing in practice. In particular, it turns out that the rough Heston model is neither a semimartingale nor a Markov process. One remedy to this problem is to use Markovian approximations of the process $V$. This essentially amounts to replacing the kernel $K$ by an approximation
\begin{equation}\label{eqn:KNForm}
K^N(t) \coloneqq \sum_{i=1}^N w_i e^{-x_i t},
\end{equation}
yielding a Markovian approximation $(S^N, \bm{V}^N)$ of $(S, V)$, where $\bm{V}^N$ is an $N$-dimensional diffusion, see e.g. \cite{abi2019markovian, abi2019multifactor, abi2019affine, alfonsi2021approximation, bayer2023markovian, bayer2023++Weak, harms2019strong}. Indeed, denote $\bw \coloneqq (w_i)_{i=1}^N$ and $\bm{V}^N \coloneqq (V^{(i)})_{i=1}^N$, and let $\bm{v}_0 = (\bm{v}_0)_{i=1}^N$ be any vector of initial conditions satisfying $\bm{w}\cdot \bm{v}_0 = V_0.$ Then, the dynamics of the Markovian approximation $(S^N, \bm{V}^N)$ are given by 
\begin{align}
\dd S^N_t &= \sqrt{V^N_t} S^N_t \left(\rho \sdd W_t + \sqrt{1-\rho^2} \sdd B_t\right),\quad S_0 = S_0, \label{eqn:RHestonMarkovStock}\\
\dd V^{(i)}_t &= -x_i(V^{(i)}_t-v^i_0)\sdd t + (\theta-\lambda V^N_t)\sdd t + \nu\sqrt{V^N_t}\sdd W_t,\quad V^{(i)}_0 = v^i_0, \label{eqn:RHestonMarkovVol}
\end{align}
for $i=1,\dots,N$, where $V^N_t = \bm{w}\cdot \bm{V}^N_t$. We remark that for dimension $N=1$, we again obtain the standard Heston model \eqref{eqn:HestonStock}-\eqref{eqn:HestonVol}.

Given these Markovian approximations, it is natural to ask for an error bound for the error between $S$ and $S^N$, and furthermore, how the nodes $\bm x\coloneqq (x_i)_{i=1}^N$ and weights $\bm w$ should be chosen in practice. This question was answered in the recent work \cite{bayer2023++Weak}, where the authors showed that for sufficiently nice payoff functions $f:\R_+\to\R$, we have the weak error bound
\begin{equation}\label{eqn:WeakErrorBoundIntroduction}
\abs{\E f(S_T) - \E f(S^N_T)} \le C\int_0^T \abs{K(t) - K^N(t)} \sdd t.
\end{equation}
This weak error bound was obtained by extending previous results of \cite{abi2019multifactor} on the characteristic function, which is known semi-explicitly for both rough Heston itself (cf. \cite{el2019characteristic}), and for its Markovian approximation (cf. \cite{abi2019markovian, abi2019affine}).

It then seems reasonable to choose the Markovian approximation $K^N$, i.e. the nodes $\bm x$ and weights $\bm w$ such that the error bound \eqref{eqn:WeakErrorBoundIntroduction} is minimized. Indeed, the same authors showed in \cite{bayer2023++Weak} that using Gaussian quadrature rules, we can achieve the rate of convergence $$\int_0^T \abs{K(t) - K^N(t)} \sdd t \le C\exp\left(-2.38\sqrt{(H + 1/2)N}\right).$$ They additionally gave an algorithm called ``BL2'' which seems to vastly outperform the Gaussian quadrature rules at least for small to moderate dimension $N$. Indeed, using the approximations $K^N$ given by ``BL2'' they manage to achieve relative errors in implied volatility smiles of well below $1\%$ for only $N=2$, yielding rather low-dimensional highly accurate Markovian approximations of $V$.

These optimized quadrature rules of \cite{bayer2023++Weak} allow us to efficiently price several options using Fourier inversion techniques. Indeed, due to the Markovian structure of the process $(S^N, \bm V^N)$, Fourier inversion using this approximation is usually much faster than applying Fourier inversion directly for $(S, V)$, see again \cite{bayer2023++Weak}. However, there are many options important in practice that cannot easily be priced efficiently using Fourier inversion, especially path-dependent options like Bermudan or American options. For such pricing tasks, it may be more beneficial to use Monte Carlo (MC) or quasi Monte Carlo (QMC) simulation. 

However, simulation of rough Heston, and even standard Heston itself, is notoriously difficult, see for example \cite{alfonsi2015affine, lord2010comparison}. Let us hence first give a short overview of some simulation schemes for the standard Heston model \eqref{eqn:HestonStock}-\eqref{eqn:HestonVol}. The main difficulty here arises from the square-root term $\sqrt{\widetilde V_t}$ in \eqref{eqn:HestonVol}. If one were to naively apply the Euler-Maruyama scheme to \eqref{eqn:HestonStock}-\eqref{eqn:HestonVol}, there is a non-zero probability that the simulated volatility process may become negative. Even worse, while the actual process $\widetilde V$ never becomes negative, it hits 0 with positive probability if the so-called Feller condition $2\theta > \nu$ is violated, see e.g. \cite{lamberton2011introduction}. And indeed, one can observe numerically that the Euler-Maruyama approximation of $\widetilde V$ consistently assumes negative values, especially in the regime $2\theta \le \nu$.

Of course, the problem of negative $\widetilde V$ can easily be resolved, for example by replacing $\sqrt{\widetilde V}$ by either $\sqrt{\widetilde V^+}$, with $x^+ \coloneqq x \lor 0$, or $\sqrt{|\widetilde V|}$, see also \cite{deelstra1998convergence, diop2003discretisation}. While these simulation schemes converge as the number of time steps approaches infinity, they do so rather slowly (or with a large leading constant in the error term), since $\widetilde V$ spends a lot (if not most) of its time close to the singularity at $\widetilde V = 0$.

Other, more promising methods for simulating the standard Heston model are based on moment-matching. In this direction, we would like to particularly highlight the QE scheme of \cite{andersen2007efficient}, and the moment-matching scheme of \cite{lileika2021second}. The QE scheme (where QE stands for quadratic-exponential) essentially approximates $\widetilde V_{t+h}$ given $\widetilde V_t$ by using either a squared Gaussian random variable, or a combination of an exponentially distributed and a Dirac random variable, where in either case the parameters of the squared Gaussian and exponential random variables are chosen such that the distribution of $\widetilde V_{t+h}$ is well-approximated. Whether in a specific simulation step we use the Gaussian or the exponential random variable essentially depends on how close $\widetilde V_t$ is to $0$. Since we are using squared Gaussians and exponential random variables, the non-negativity of our simulated paths $\widetilde V$ is automatically ensured. While the authors of \cite{andersen2007efficient} could not derive a rate of convergence of the QE scheme as the number of time steps approaches infinity, numerical tests seemed to indicate faster than linear rates of convergence.

Conversely, the moment-matching scheme of \cite{lileika2021second} essentially approximates $\widetilde V_{t+h}$ given $\widetilde V_t$ by using a discrete random variable that can assume three different values (and all three values are of course non-negative). This discrete random variable is chosen such that the first five moments of $\widetilde V_{t+h}$ are fitted exactly. A convenient advantage of this scheme over the QE scheme is that no regime-switching between two different kinds of approximation occurs. Moreover, they manage to show that this moment-matching scheme converges with second order, at least for the volatility process $\widetilde V$.

When trying to simulate samples of the rough Heston model, of course similar problems with the square root and the non-negativitiy of the volatility process occur, which is why a simple Euler scheme will not yield good convergence results. A further problem now enters due to the lack of Markov property. If one wanted to apply Euler directly to \eqref{eqn:RHestonStock}-\eqref{eqn:RHestonMarkovVol}, one would additionally have to compute the integrals in \eqref{eqn:RHestonVol} in every time step. This method would thus have quadratic computational cost.

Indeed, the authors of this paper are aware of only one viable simulation scheme, which is the HQE scheme of \cite{gatheral2022efficient}. The HQE scheme is essentially an adaptation of the QE scheme for rough Heston or more general affine rough forward rate models. While the HQE scheme numerically exhibits a weak rate of convergence of 1 in the number of time discretization steps, its computational cost scales quadratically, precisely because of the need to compute the integrals in \eqref{eqn:RHestonMarkovVol}.

The aim of this paper is to leverage the highly accurate low-dimensional Markovian approximations of \cite{bayer2023++Weak} to overcome the quadratic cost in the number of time steps. Hence, in Section \ref{sec:MackeviciusScheme}, we extend the weak simulation scheme of \cite{lileika2021second} based on moment-matching with a discrete-valued random variable to the setting of the Markovian approximations \eqref{eqn:RHestonMarkovStock}-\eqref{eqn:RHestonMarkovVol}. Then, in Section \ref{sec:Numerics}, we compare this weak Markovian scheme with an Euler scheme based on the same Markovian approximation (henceforth denoted as \emph{Markovian Euler scheme} to avoid confusion with a standard Euler discretization of rough Heston), and with the HQE scheme of \cite{gatheral2022efficient}. In particular, we compare the efficiency of these three simulation schemes for computing implied volatility smiles, implied volatility surfaces, Asian call option prices, and Bermudan put option prices in Section \ref{sec:NumericalResults}.

In all our numerical examples, our weak Markovian scheme significantly outperforms both the HQE scheme and the Markovian Euler scheme. 
While we were unable to prove convergence of our weak scheme (see Section \ref{sec:WellDefinitenessAndConvergence}) at this point, extensive numerical experiments suggests that the error decays like $O(n^{-2})$, where $n$ is the number of time discretization steps, i.e., convergence with rate two. For comparison, both the Markovian Euler scheme and the HQE scheme exhibit numerical convergence at most with rate one. 
Another advantage of the Markovian schemes is that they have a computational cost of $O(n)$, compared to $O(n^2)$ for the HQE scheme. Indeed, the HQE scheme needs to carry along the past values of the process, and recompute the integrals in \eqref{eqn:RHestonVol} in every time step. A further advantage of the Markovian schemes over the HQE scheme becomes apparent in the pricing of Bermudan options: Since we have Markov processes, we may simply apply standard pricing schemes like the Longstaff-Schwartz algorithm, while accurate pricing under the HQE scheme requires either more sophisticated (non-Markovian) pricing schemes, or the inclusion of past values in the Longstaff-Schwartz algorithm.

\section{Weak simulation scheme}\label{sec:MackeviciusScheme}

\subsection{Derivation of the scheme}\label{sec:DerivationOfWeakScheme}

Looking at the dynamics of the Markovian approximation \eqref{eqn:RHestonMarkovStock}-\eqref{eqn:RHestonMarkovVol}, it is immediately clear that the volatility process $\bm{V}$ does not depend on the stock price $S$. Hence, we will first give a scheme for simulating $\bm{V}$, and later a scheme for simulating $S$ while preserving the correlation $\rho$ between the Brownian motions driving $\bm{V}$ and $S$.

\subsubsection{Simulating the volatility}\label{sec:VolatilityAlgorithm}

Consider first the Markovian approximation of the volatility process only. Recall that we have 
\begin{equation*}
\dd V^{(i)}_t = -x_i(V^{(i)}_t-v^i_0)\sdd t + (\theta-\lambda V^N_t)\sdd t + \nu\sqrt{V^N_t}\sdd W_t,\ V^{(i)}_0 = v^i_0,\ i=1,\dots,N,
\end{equation*}
where $\bw \coloneqq (w_i)_{i=1}^N$ is the vector of weights, $\bm{v}_0 = (\bm{v}_0)_{i=1}^N$ is any initial condition with $\bw \cdot \bv_0 = V_0$, $\bm{V}^N \coloneqq (V^{(i)})_{i=1}^N$, and $V^N_t = \bm{w}\cdot \bm{V}^N_t$.

In \cite{lileika2021second}, the authors studied weak approximations using discrete-valued random variables of the CIR process, and showed that their approximations converge weakly with second order. The dynamics of $\bm{V}^N$ are quite similar in nature to the dynamics of the CIR process, and indeed, for $N=1$ we obtain the CIR process. Thus, we aim to give a simulation scheme for $\bm{V}^N$ similar to that in \cite{lileika2021second}.

Similarly as in \cite{lileika2021second}, we split the above SDE in two parts. We denote by $D(\bm{z}, h) \coloneqq \bm{Z}_h \coloneqq Z^i_h$, the solution at time $h$ of the ODE $$\dd Z^i_t = -x_i(Z^i_t - v^i_0)\sdd t + (\theta - \lambda Z_t) \sdd t,\quad Z^i_0 = z^i,\quad i=1,\dots,N,\quad Z_t = \bm{w} \cdot \bm{Z}_t,$$ and by $S(\bm{y}, h) \coloneqq \bm{Y}_h \coloneqq Y^i_h$ the solution at time $h$ of the SDE
\begin{equation}\label{eqn:VMarkovStochasticPart}
\dd Y^i_t = \nu\sqrt{Y_t}\sdd W_t,\qquad Y^i_0 = y^i,\qquad i=1,\dots, N,\qquad Y_t = \bm{w} \cdot \bm{Y}_t.
\end{equation}

Next, we give schemes $\widehat{D}$ and $\widehat{S}$ for approximating $D$ and $S$, respectively. First, note that $D$ is a linear ODE and can hence be solved exactly. The solution is given by $$D(\bz, h) = \bm{Z}_h = e^{Ah}\bm{z} + A^{-1} (e^{A h} - \id) b,$$ where $$A \coloneqq -\lambda \bm{1}\bm{w}^T - \diag(\bm{x}),\quad\text{and}\quad b \coloneqq \theta\bm 1 + \diag(\bm{x}) \bm{v}_0,$$  where $\bm 1 \coloneqq (1, \dots, 1)^T\in\R^N$, $\id\in\R^{N\times N}$ is the identity matrix, and $\diag(\bm x)\in\R^{N\times N}$ is the diagonal matrix with entries $\bm x$. Therefore, we can simply set $\widehat{D}(\bz, h) \coloneqq D(\bz, h)$.

Next, we provide a scheme $\widehat{S}$ for approximating $S$. Define $\overline{w} \coloneqq \bm{1}^T \bm{w}$. Then, we may take the inner product of the SDE \eqref{eqn:VMarkovStochasticPart} with $\bm{w}$ to get the SDE $$\dd Y_t = \nu\overline{w}\sqrt{Y_t} \sdd W_t,\qquad Y_0 = \bm{w}\cdot \bm{y}.$$ This is exactly the SDE studied in \cite{lileika2021second}, with $x \coloneqq \bm{w}\cdot\bm{y}$ and $z\coloneqq\nu^2\overline{w}^2 h.$ They give the following second order scheme. Define the quantities
\begin{align*}
m_1 &= x,\qquad m_2 = x^2 + xz,\qquad m_3 = x^3 + 3x^2z + \frac{3}{2}xz^2,\\
p_1 &= \frac{m_1x_2x_3 - m_2(x_2 + x_3) + m_3}{x_1(x_3-x_1)(x_2-x_1)},\\
p_2 &= \frac{m_1x_1x_3 - m_2(x_1 + x_3) + m_3}{x_2(x_3-x_2)(x_1-x_2)},\\
p_3 &= \frac{m_1x_1x_2 - m_2(x_1 + x_2) + m_3}{x_3(x_1-x_3)(x_2-x_3)},\\
x_1 &= x + Az - \sqrt{\left(3x + A^2z\right) z},\\
x_2 &= x + \left(A-\frac{3}{4}\right) z,\\
x_3 &= x + Az + \sqrt{\left(3x + A^2z\right) z},\\
A &= \frac{6 + \sqrt{3}}{4}.
\end{align*}
Then, we define $\widehat Y_h$ to be the random variable which is $x_i$ with probability $p_i$, $i=1,2,3$. We remark that for $i=1,2,3$, $m^i = \E Y_h^i$ and $x_i \ge 0$, and that the $x_i$ and $p_i$ are chosen such that $p_1+p_2+p_3=1$ and $\E \widehat Y_h^k = \E Y_h^k$ for $k=1,\dots,5$.

But now, we are not merely interested in simulating $Y$, but actually in simulating the entire process $\bm{Y}$. Fortunately, we can easily reconstruct an approximation $\widehat{\bm{Y}}$ from $\widehat Y$. Indeed, we note that the right-hand side of \eqref{eqn:VMarkovStochasticPart} is the same for all $i=1,\dots,N$. Hence, the solution of \eqref{eqn:VMarkovStochasticPart} must be of the form
\begin{equation}\label{eqn:BYForm}
Y^i_h = y^i + Q,\qquad i = 1,\dots,N,
\end{equation}
for some scalar random variable $Q$. Taking the inner product of \eqref{eqn:BYForm} with $\bm{w}$, we get $$Y_h = \bw\cdot\by + \overline{w}Q,\quad \textup{implying}\quad Q = \frac{Y_h - \bw\cdot\by}{\overline{w}}.$$ Hence, we set $$\widehat{S}(\by, h) \coloneqq \widehat{\bm{Y}}_h \coloneqq \by + \frac{\widehat Y_h - \bw\cdot\by}{\overline{w}}.$$

Finally, we combine the schemes for drift and diffusion by Strang splitting, and get 
\begin{equation}\label{eqn:ACIR}
A^{\textup{CIR}}(\bm{v},h) \coloneqq D\left(\widehat{S}\left(D\left(\bm{v}, \frac{h}{2}\right), h\right), \frac{h}{2}\right)
\end{equation}
for approximating $\bm{V}_h$ given $\bm{v}$.

\subsubsection{Simulating the stock price}\label{sec:StockPriceAlgorithm}

We now have a scheme $A^{\textup{CIR}}$ for simulating the volatility process $\bm{V}$. Next, we give a scheme for simulating $(S, \bm{V})$ as in \eqref{eqn:RHestonMarkovStock}-\eqref{eqn:RHestonMarkovVol}. The difficulty here is that the Brownian motion in the stock price is correlated with the Brownian motion driving the volatility, and the scheme $A^{\textup{CIR}}$ does not require the simulation of increments of $W$.

The ensure that we have the correct correlation between $S$ and $\bm{V}$, we follow \cite{alfonsi2010high}. Define the processes $$Y^i_t = \int_0^t V^i_t \sdd t,\qquad i=1,\dots,N.$$ The process $\bm Y \coloneqq (Y^i)_{i=1}^N$ will be useful, since the solution $S$ to \eqref{eqn:RHestonMarkovStock} is given by $$S_t = S_0 \exp\left(\int_0^t \sqrt{V_s}\left(\rho \sdd W_s + \sqrt{1-\rho^2}\sdd B_2\right) - \frac{1}{2}\int_0^t V_s \sdd s\right),$$ where the latter term in the exponent is precisely $-\frac{1}{2}\bm w \cdot \bm Y$.

Now, the SDE $S^{\textup{rHeston}}((s, \bm{v}, \bm{y}), h)$ given by 
\begin{align*}
\dd S_t &= \sqrt{V_t}S_t\left(\rho \sdd W_t + \sqrt{1-\rho^2} \sdd B_t\right),\qquad S_0=s,\\
\dd V^i_t &= -x_i(V^i_t - v^i_0)\sdd t + (\theta - \lambda V_t) \sdd t + \nu \sqrt{V_t} \sdd W_t,\qquad V^i_t = v^i,\\
\dd Y^i_t &= V^i_t \sdd t,\qquad Y^i_0=y^i,
\end{align*}
for $i=1,\dots,N$ can again be split into two parts, namely the SDE $S^W((s, \bm{v}, \bm{y}), h)$ given by 
\begin{align*}
\dd S_t &= \sqrt{V_t}S_t\rho \sdd W_t,\qquad S_0=s,\\
\dd V^i_t &= -x_i(V^i_t - v^i_0)\sdd t + (\theta - \lambda V_t) \sdd t + \nu \sqrt{V_t} \sdd W_t,\qquad V^i_t = v^i,\\
\dd Y^i_t &= V_t \sdd t,\qquad Y^i_0=y^i,
\end{align*}
and the SDE $S^B((s,\bm{v},\bm{y}), h)$ given by
\begin{align*}
\dd S_t &= \sqrt{V_t}S_t\sqrt{1-\rho^2} \sdd B_t,\qquad S_0=s,\\
\dd V^i_t &= 0,\qquad V^i_t = v^i,\\
\dd Y^i_t &= 0,\qquad Y^i_0=y^i.
\end{align*}

The SDE $S^B$ can be solved explicitly. Indeed, since the volatility $V$ is constant, this is essentially the Black-Scholes model. The solution is given by 
\begin{align*}
S_h &= s\exp\left(\sqrt{v}\sqrt{1-\rho^2}B_h - \frac{1}{2}v(1-\rho^2) h\right),\\
\bm{V}_h &= \bm{v},\\
\bm{Y}_h &= \bm{y}.
\end{align*}

For the SDE $S^W$, we first approximate $\widehat{\bm V}_h \coloneqq A^{\textup{CIR}}(\bm v, h).$ Next, we approximate $\bm Y$ using the trapezoidal rule, i.e. $\widehat{\bm Y}_h \coloneqq \bm y + (\bm v + \widehat{\bm V}_h)/2$. Next, we want to express $S_h$ in terms of $\bm{V}_h$ and $\bm{Y}_h$. We make the Ansatz
\begin{align}
S_t &= s\exp\left(at + \sum_{i=1}^N b_i(Y^i_t - Y^i_0) + \sum_{i=1}^N c_i(V^i_t - V^i_0)\right).\label{eqn:Ansatz}
\end{align}

Using the Itô formula, we get
\begin{align*}
dS_t &= S_t\left[a\sdd t + \sum_{i=1}^N b_i \sdd Y^i_t + \sum_{i=1}^N c_i \sdd V^i_t + \frac{1}{2}\sum_{i,j=1}^N c_ic_j\sdd [V^i,V^j]_t\right]\\
&= S_t\Bigg[\left(a + \sum_{i=1}^N c_ix_iv^i_0 + \theta \sum_{i=1}^N c_i\right)\sdd t\\
&\qquad + \sum_{i=1}^N \left(b_i - c_ix_i - \lambda w_i\sum_{j=1}^N c_j + \frac{1}{2}\nu^2 w_i\sum_{j,k=1}^N c_jc_k\right) V_i \sdd t + \nu\sum_{i=1}^N c_i\sqrt{V_t} \sdd W_t\Bigg].
\end{align*}

To retain the SDE $S^W$, we thus require that 
\begin{align*}
a + \sum_{i=1}^N c_ix_iv^i_0 + \theta \sum_{i=1}^N c_i &= 0,\\
b_i - c_ix_i - \lambda w_i\sum_{j=1}^N c_j + \frac{1}{2}\nu^2 w_i\sum_{j,k=1}^N c_jc_k &= 0,\qquad i=1,\dots,N,\\
\nu\sum_{i=1}^N c_i &= \rho.
\end{align*}

This is achieved with
\begin{align*}
a &= -\sum_{i=1}^N c_ix_iv^i_0 - \theta \sum_{i=1}^N c_i,\\
b_i &= c_ix_i + \lambda w_i\sum_{j=1}^N c_j - \frac{1}{2}\nu^2 w_i\sum_{j,k=1}^N c_jc_k,\qquad i=1,\dots,N,\\
\sum_{i=1}^N c_i &= \frac{\rho}{\nu},
\end{align*}
where the last equation of course has many solutions. We may thus choose the $(c_i)$ to be such that the representation \eqref{eqn:Ansatz} (at least heuristically) becomes as numerically stable as possible. Assume without loss of generality that the nodes $(x_i)$ are ordered with $0 \le x_1 < \dots < x_N$. Then, we choose
\begin{equation}\label{eqn:ChoiceOfCi}
c_1 = \frac{\rho}{\nu},\qquad c_i = 0,\qquad i = 2,\dots, N.
\end{equation}
This further yields
\begin{align*}
a &= - \left(x_1 v^1_0 + \theta \right)\frac{\rho}{\nu},\qquad b_i = \frac{\rho}{\nu} x_1\delta_{i,1} + \lambda w_i \frac{\rho}{\nu} - \frac{1}{2} w_i \rho^2.
\end{align*}
We remark that we have tried several different choices of $(c_i)$, and \eqref{eqn:ChoiceOfCi} seemed to perform best, especially for larger mean-reversions $\bm x$. Therefore, we get the solution
\begin{align*}
S_t &= s\exp\bigg(\frac{\rho}{\nu}\bigg(- \left(x_1 v^1_0 + \theta \right)t + x_1 (Y^1_t - Y^1_0)\\
&\qquad + \left(\lambda - \frac{1}{2}\rho\nu\right)(Y_t - Y_0) + (V^1_t - V^1_0)\bigg)\bigg),
\end{align*}
where $Y_t \coloneqq \bm{w}\cdot \bm{Y}$.

Hence, we may solve the SDE $S^W$ approximately with the algorithm $\widehat{S}^W$ given by
\begin{align*}
\widehat{\bm{V}}_h &= A^{\textup{CIR}}(\bm{v}, h),\\
\widehat{\bm{Y}}_h &= \bm{y} + (\bm{v} + \widehat{\bm{V}}_h)\frac{h}{2},\\
\widehat S_h &= s\exp\bigg(\frac{\rho}{\nu}\bigg(- \left(x_1 v^1_0 + \theta \right)h + x_1 (\widehat Y^1_h - Y^1_0)\\
&\qquad + \left(\lambda - \frac{1}{2}\rho\nu\right)(\widehat Y_h - Y_0) + (\widehat V^1_h - V^1_0)\bigg)\bigg).
\end{align*}

We have thus solved the SDEs $S^B$ and $S^W$ separately, and now want to combine them again using some splitting scheme. While the algorithm for $S^B$ is exact, the algorithm for $\widehat S^W$ may be of second order, since $A^{\textup{CIR}}$ is based on the second-order algorithm of \cite{lileika2021second}, and $\widehat{\bm Y}$ is computed using the trapezoidal rule. Hence, we would like to use a splitting scheme that is also of second order. We remark that there are several such splitting schemes, see for example the overview article \cite{mclachlan2002splitting}. One possibility is to again use Strang splitting, as in \eqref{eqn:ACIR} for $A^{\textup{CIR}}$. However, in our experience the following randomized Leapfrog splitting seemed to perform slightly better. This splitting scheme is given by $$\widehat{S}^{\textup{rHeston}}((s,\bm{v},\bm{y}), h) =
\begin{cases}
\widehat{S}^W(S^B((s,\bm{v},\bm{y}), h), h),\qquad &\textup{if } U \le \frac{1}{2},\\
S^B(\widehat{S}^W((s,\bm{v},\bm{y}), h), h),\qquad &\textup{else},
\end{cases}$$ where $U$ is a uniform random variable on $[0, 1]$ independent of everything else.

\subsection{Boundary behaviour and convergence}\label{sec:WellDefinitenessAndConvergence}

In Section \ref{sec:DerivationOfWeakScheme} we have given an algorithm for simulating the Markovian approximation of the rough Heston model \eqref{eqn:RHestonMarkovStock}-\eqref{eqn:RHestonMarkovVol}. Let us now first discuss whether the algorithm is well-defined, i.e. whether the total volatility $V$ stays non-negative. The only apparent issue that could occur is that the scheme $A^{\textup{CIR}}$ for simulating the volatility $\bm{V}$ may yield negative values for $V = \bw\cdot \bm{V}$. Here, we recall that $A^{\textup{CIR}}$ was given by $$A^{\textup{CIR}}(\bv, h) = D\left(\widehat{S}\left(D\left(\bv, \frac{h}{2}\right),h\right),\frac{h}{2}\right),$$ where $D$ was the (exact) solution to an ODE and $\widehat{S}$ was the approximate solution to an SDE.

Recall also that the algorithm $\widehat{S}$ is a simple multidimensional extension of \cite{lileika2021second}, who studied the one-dimensional equation
\begin{equation*}
\dd X_t = \alpha \sqrt{X_t} \sdd W_t.
\end{equation*}
The authors in \cite{lileika2021second} actually proved that their scheme for simulating $X$ remains non-negative. In particular, this implies that $\widehat{V}_h \coloneqq \widehat{S}(\bv, h)$ satisfies $\bw\cdot \widehat{V}_h \ge 0$ if $\bw\cdot \bv \ge 0$. Therefore, the step $\widehat{S}$ always retains non-negativity.

\begin{figure}
\centering
\includegraphics[scale=0.6]{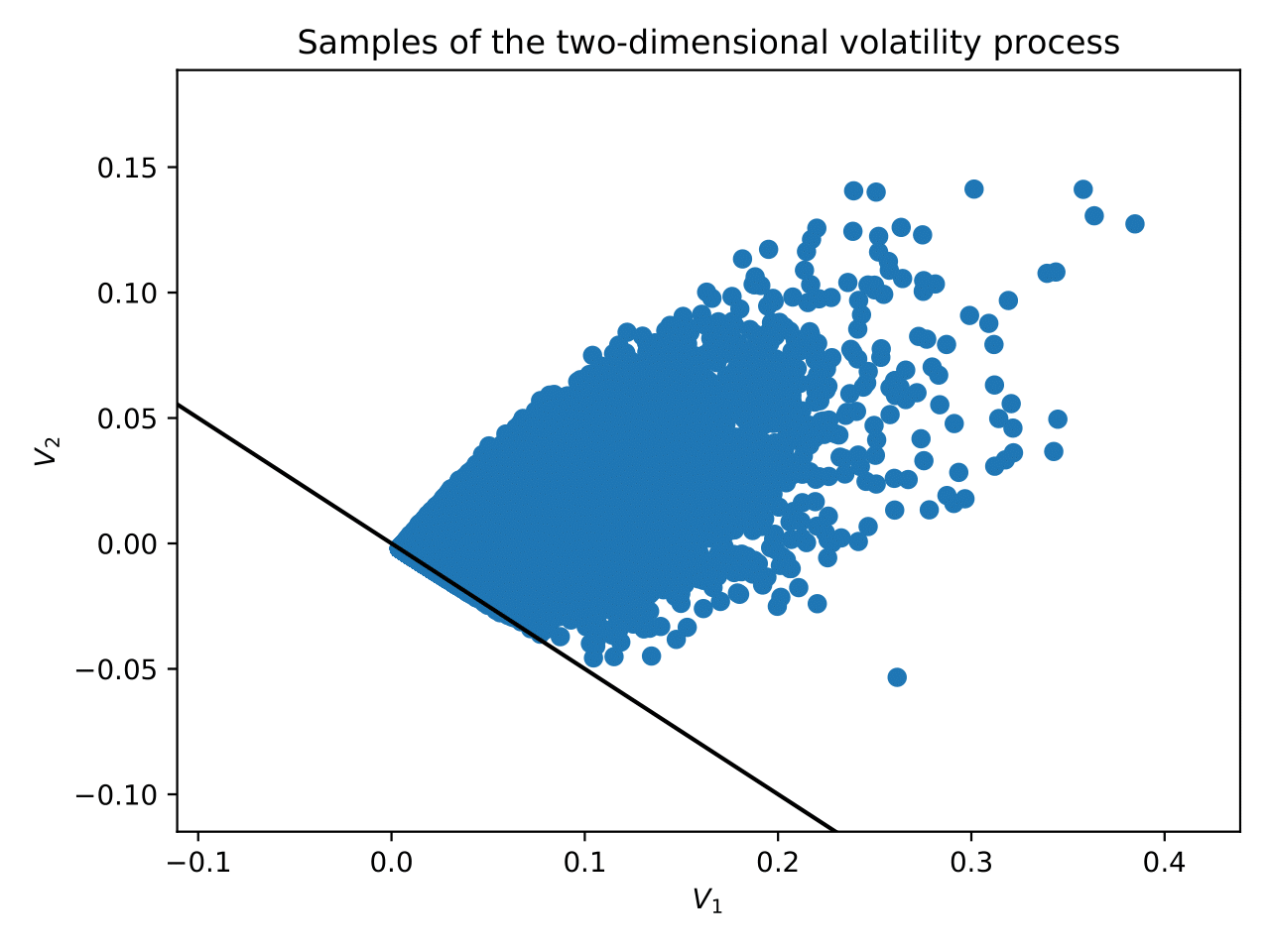}
\caption{Samples of the two-dimensional Markovian approximation $\bm{V}^{N, M}$ of the volatility using $10^5$ sample paths on a time grid with $1000$ time steps. Plotted are all the points $\bm{V}^{N,M}_{t_i}$ for every time step $t_i$, $i=0,\dots,1000$, and all the $10^5$ samples. The black line is the line where the total volatility $\bw\cdot \bm{V}^{N,M} = 0$, and the total volatility is positive above that line. We can see clearly that $\bm{V}^{N, M}$ seems to lie in a cone in the non-negative half-space. If it seems like $\bm{V}^{N, M}$ occasionally lies below the black line, this is merely because of the thickness of the sample points, zooming in shows that all samples are always above the black line. The parameters used are $\bm{x} = (1, 10), \bw=(1, 2), \lambda=0.3,\nu=0.3,V_0=0.02,\theta=0.02,T=1,\bm{v}_0=(0.02, 0).$}
\label{fig:SamplesOfTwoDimensionalVolatilityProcess}
\end{figure}

The situation is a lot less clear for the deterministic part $D$. Indeed, it is not hard to find two-dimensional examples where $\bw\cdot \bv\ge 0$, but $\bw\cdot D(\bv, h) < 0$. However, in our numerical simulations it never happened that the algorithm actually produced negative volatility. In fact, it seems that the process $\bm{V}$ does not only stay in the half-space $\bw\cdot\bm{V} \ge 0$, but even in a cone that is contained in it. Unfortunately we were unable to prove that this is really the case, but in Figure \ref{fig:SamplesOfTwoDimensionalVolatilityProcess} we provide a plot illustrating that point. However, together with Eduardo Abi Jaber we are currently working on resolving this issue and determining the state space of the Markovian approximation $\bm V$, see \cite{bayer2023+state}.

While it therefore seems that $A^{\textup{CIR}}$ produces only non-negative total volatilities, we were not able to prove that. Hence, the algorithm above is strictly speaking not well-defined. To remedy this, one could for example use $V^+ \coloneqq \max(V, 0)$ instead of $V$ whenever non-negativity is crucial. However, since this never seems to occur we have in fact not done so in our implementation.

The next question is about the convergence of the approximation scheme. As noted above already, the authors in \cite{lileika2021second} proved weak second order convergence for their approximation scheme of the CIR process. In particular, this implies that $\bm{V}^{N, M}\to \bm{V}^N$ weakly of order 2 for $N=1$, since for $N=1$ the process $V^N$ is the CIR process. Since $\bm{V}^N$ for $N\ge 2$ seems to be merely a multidimensional version of $\bm{V}^1$, we conjecture that $\bm{V}^{N,M}\to \bm{V}^N$ also weakly of order 2 for any $N$. However, since we were not even able to prove that $\bw\cdot \bm{V}^{N, M}$ stays non-negative, we could of course also not prove a convergence rate.

Next, we used an algorithm very similar to the one in \cite[Section 4.2]{alfonsi2010high} for approximating $S$ given an approximation of $\bm{V}^N$. While the scheme in \cite[Section 4.2]{alfonsi2010high} is modelled after second-order schemes in \cite[Section 1]{alfonsi2010high}, the author notes that it is not directly possible to prove a convergence rate of order two, since the proofs in \cite[Section 1]{alfonsi2010high} require that the SDE has uniformly bounded moments. For the Heston model this is not necessarily the case, as was shown in \cite{gerhold2019moment}. Similar comments of course apply in our setting.

\section{Numerics}\label{sec:Numerics}

In this section we price various kinds of vanilla and exotic options in the rough Heston model, where we compare the weak approximation scheme from Section \ref{sec:MackeviciusScheme} with a simple implicit Euler scheme given in Section \ref{sec:ImplicitEulerScheme} and the HQE scheme from \cite{gatheral2022efficient} as in Section \ref{sec:HQEScheme}. In all our examples we used the same parameters as in \cite[Section 4.2]{abi2019multifactor}, namely $$\lambda=0.3,\ \nu=0.3,\ \theta=0.02,\ V_0=0.02,\ \rho=-0.7,\ S_0=1,$$ while the values for $T$ and $H$ may vary. We remark that the weak scheme and the implicit Euler scheme require us to choose a Markovian approximation $K^N$ of $K$ of the form \eqref{eqn:KNForm}. The recent paper \cite{bayer2023++Weak} studies the problem of weak Markovian approximation of rough Heston, and the algorithm ``BL2'' shows promising results. Hence, we decided to use the approximations $K^N$ generated by this algorithm BL2. Throughout, we will explicitly write down the nodes $\bx$ and weights $\bw$ that were generated using BL2.

This section is therefore structured as follows. First, in Section \ref{sec:OtherSimulationSchemes}, we briefly describe the implicit Euler scheme in Section \ref{sec:ImplicitEulerScheme} and the HQE scheme in Section \ref{sec:HQEScheme} that we compare our weak scheme with. Furthermore, in Section \ref{sec:SimulationCost}, we compare the computational costs of these three simulation schemes. Finally, in Section \ref{sec:NumericalResults} we compare these three simulation schemes for various kinds of options: Implied volatility smiles in Section \ref{sec:EuropeanCallSmiles}, implied volatility surfaces in Section \ref{sec:EuropeanCallSurfaces}, Geometric Asian options in \ref{sec:GeometricAsian}, and Bermudan put options in \ref{sec:BermudanPut}.

\subsection{Other simulation schemes used for benchmarking}\label{sec:OtherSimulationSchemes}

\subsubsection{Implicit Euler scheme}\label{sec:ImplicitEulerScheme}

Since the Markovian approximation $(S^N, \bm{V}^N)$ given in \eqref{eqn:RHestonMarkovStock}-\eqref{eqn:RHestonMarkovVol} satisfies a standard SDE, we can apply a (slightly modified) implicit Euler scheme to simulate sample paths. We use the implicit Euler scheme rather than the explicit Euler scheme due to the stiffness of the SDE governing $\bm{V}^N$. For large speeds of mean-reversion $\bm x$, the explicit Euler scheme tends to overshoot, leading to highly unstable behaviour for $\Delta t \gtrsim x_N^{-1}$. The implicit Euler scheme, however, mostly suppresses those large oscillations, avoiding instability and overflow errors.

Due to the square root in the dynamics, we have to ensure that $V^{N,M}$ stays non-negative. There are various ways to do this, and we have arguably chosen the simplest one. We simply truncate $V^{N,M}$ by taking the positive part. Another simple solution to this problem may be for example reflection, i.e. using $\abs{V^{N,M}}$ instead of $(V^{N,M})^+$. Overall, the quality of the simulation results does not change much depending on the specific way one deals with this issue.

Choose a final time $T$, and a number of steps $M$. We discretize the interval $[0,T]$ into $M$ subintervals of equal length $\Delta t \coloneqq T/M$. Hence, we obtain a partition $(t_m)_{m=0}^M$ of $[0,T]$ with $t_m = m\Delta t$. On this time grid, we define the approximation $(S^{N,M}, \bm{V}^{N,M})$ of $(S^N,\bm{V}^N)$ using the drift-implicit Euler scheme given by
\begin{align*}
S^{N,M}_{t_{m+1}} &= S^{N,M}_{t_m} \sqrt{(V^{N,M}_{t_m})^+}S^{N,M}_{t_m}\left(\rho \Delta W_m + \sqrt{1-\rho^2}\Delta B_m\right),\\
V^{(i),M}_{t_{m+1}} &= V^{(i),M}_{t_m} -x_i(V^{(i),M}_{t_{m+1}} - v^i_0)\Delta t + (\theta - \lambda V^{N,M}_{t_{m+1}})\Delta t + \nu\sqrt{(V^{N,M}_{t_m})^+}\Delta W_m,
\end{align*}
for $m=0,\dots,M-1$. Here, we set $S^{N,M}_0 = S_0$, $V^{(i),M}_0 = v^i_0$, and $\bm{V}^{N, M}_{t_m} \coloneqq (V^{(i), M}_{t_m})_{i=1}^N$. Also, $x^+ \coloneqq x\lor 0$, and we have the Brownian increments $\Delta W_m \coloneqq W_{t_{m+1}}-W_{t_m}$ and $\Delta B_m \coloneqq B_{t_{m+1}} - B_{t_m}$. We remark that we can obtain $\bm{V}_{m+1}^{N,M}$ from $\bm{V}_m^{N,M}$ by solving an $N$-dimensional linear system.

\subsubsection{HQE scheme}\label{sec:HQEScheme}

Secondly, we compare the weak scheme from Section \ref{sec:MackeviciusScheme} with the HQE scheme developed in \cite{gatheral2022efficient}, building on the classical QE scheme for the (standard) Heston model, cf. \cite{andersen2007efficient}. The HQE scheme does not rely on the Markovian approximations, but rather works directly on the rough volatility process. While a convergence rate has not been theoretically established for the HQE scheme, it seems to numerically exhibit order 1 convergence as the number of time steps goes to infinity. The HQE scheme has two major disadvantages compared to our schemes using Markovian approximations. First, due to the non-Markovianity, the computational cost scales quadratically in the number of time steps. Second, since the simulated paths lack the Markov structure, it is more difficult to price e.g. Bermudan options using Longstaff-Schwartz, see also Section \ref{sec:BermudanPut}.

\subsubsection{Computational cost of simulation}\label{sec:SimulationCost}

Let us briefly compare the computational cost for simulating paths using our weak scheme from Section \ref{sec:MackeviciusScheme}, the Euler scheme from Section \ref{sec:ImplicitEulerScheme}, and the HQE scheme from \cite{gatheral2022efficient}. Due to the non-Markovian structure, the HQE scheme has the cost $O(M^2)$ per sample, where $M$ is the number of time steps. In contrast, the cost of the weak and the Euler scheme is $O(N^2M)$, where $N$ is the dimension of the Markovian approximation. However, as we will see in Section \ref{sec:NumericalResults}, $N\in\{2,3\}$ is usually enough, so that in practice the weak and the Euler scheme exhibit linear cost. Some specific computational times are also given in Table \ref{tab:ErrorsSmileHIs01}.

\subsection{Numerical results}\label{sec:NumericalResults}

\subsubsection{European call option}\label{sec:EuropeanCallSmiles}

In our first example, we consider European call options. Since the characteristic functions of the stock price in the rough Heston model and its Markovian approximation are known in semi-closed form, we can compute reference prices using Fourier inversion, see also \cite{abi2019multifactor, abi2019affine}. Our goal is then to numerically observe convergence rates for the various discretization schemes as the number of time steps $M\to \infty$.

\begin{figure}
\centering
\includegraphics[scale=0.6]{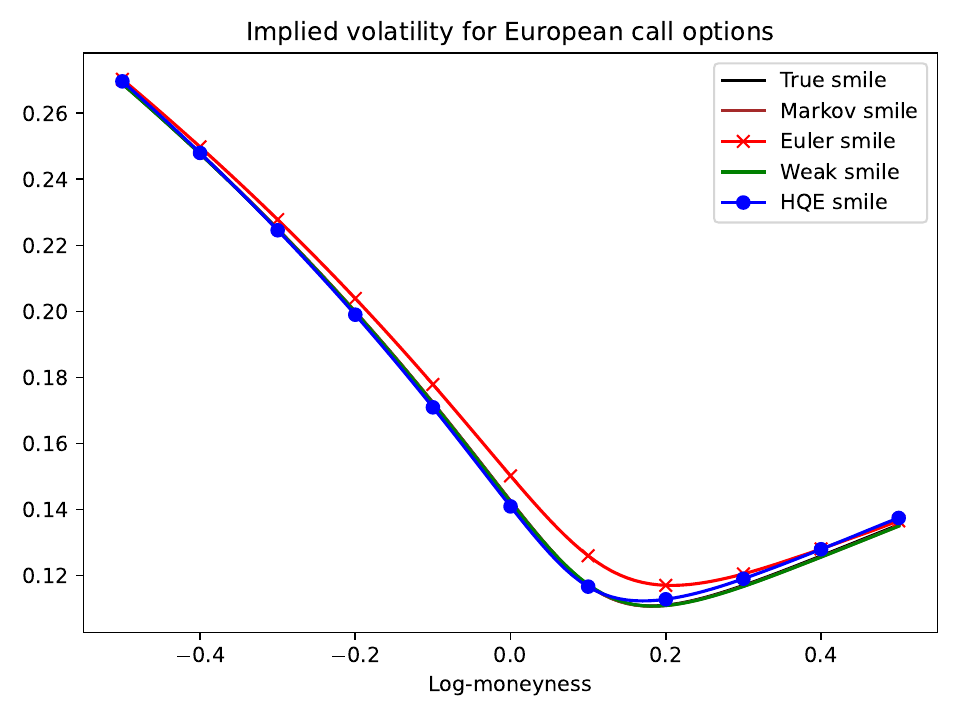}
\caption{Implied volatility smiles using $m=2^{24}$ samples and $M=32$ time steps, with $H=0.1$, $T=1$, and using the nodes and weights given in Table \ref{tab:QuadratureRuleSmile} for $N=2$. The rough Heston smile (black) and the Markovian smile (brown) obtained by Fourier inversion are not visible because they are under the weak smile (green).}
\label{fig:IVSmilesEurCall}
\end{figure}

First, in Figure \ref{fig:IVSmilesEurCall} we see the implied volatility smiles of the Euler, the weak and the HQE scheme for a fixed number of time steps $M=32$. Observe that the weak scheme is sufficiently accurate so that one cannot distinguish its smile from the truth in the plot. In contrast, the HQE scheme is still a bit off for strikes in the money, and the Euler scheme has a rather large error.

Next, we want to determine the numerical convergence rates of the implied volatility smiles. The restricting factor here is the MC error. If the MC error is too large, we will not be able to accurately determine the rate of convergence. We do two things to combat this problem. First, we restrict ourselves to strikes close to the money, since the MC error is usually smaller there. Hence, we use the maturity $T=1$ and 16 linearly spaced values in $[-0.1, 0.05]$ for the log-moneyness.

Second, we use quasi Monte Carlo (QMC) instead of MC to further reduce the error. In order to still obtain an error estimate of the QMC error similar to the MC error estimate, we use randomized QMC (RQMC), see e.g. the review article \cite{l2009quasi}. We give the number of samples generated as $m=m_1\cdot m_2$, where $m_2$ is the number of Sobol points we use, and $m_1$ is the number of random shifts of these Sobol points.

We now consider two different Hurst parameters, namely $H=0.1$ and $H=-0.2$, and for both we use $N=2$ and $N=3$ dimensions for the Markovian approximation of $V$. The precise nodes and weights used are given in Table \ref{tab:QuadratureRuleSmile}. 

\begin{table}
\centering
\begin{tabular}{c|c|c|c|c|c|c}
     && \multicolumn{2}{c|}{$N=2$} & \multicolumn{3}{c}{$N=3$}  \\ \hline
    \multirow{2}{*}{$H=0.1$} & Nodes & 0.05 & 8.7171 & 0.033333 & 2.2416 & 46.831\\ \cline{2-7}
     & Weights & 0.76733 & 3.2294 & 0.55543 & 1.1110 & 6.0858\\ \hline
    \multirow{2}{*}{$H=-0.2$} & Nodes & 0.49172 & 60.452 & 0.63781 & 9.6554 & 681.37\\ \cline{2-7}
     & Weights & 0.70202 & 33.927 & 0.66909 & 3.3694 & 184.50
\end{tabular}
\caption{Nodes and weights from the algorithm BL2 in \cite{bayer2023++Weak} for $T=1$ and $H=0.1$ (top tables), and $H=-0.2$ (bottom tables), both with $N=2$ and $N=3$.}
\label{tab:QuadratureRuleSmile}
\end{table}

\begin{figure}
\centering
\begin{minipage}{.5\textwidth}
  \centering
  \includegraphics[width=\linewidth]{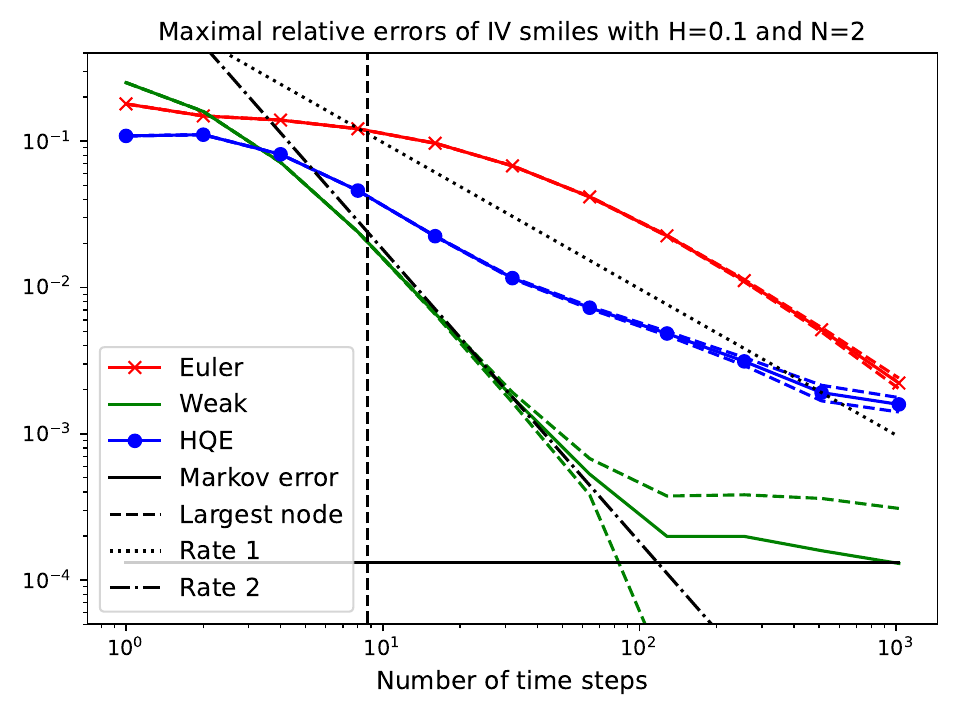}
\end{minipage}%
\begin{minipage}{.5\textwidth}
  \centering
  \includegraphics[width=\linewidth]{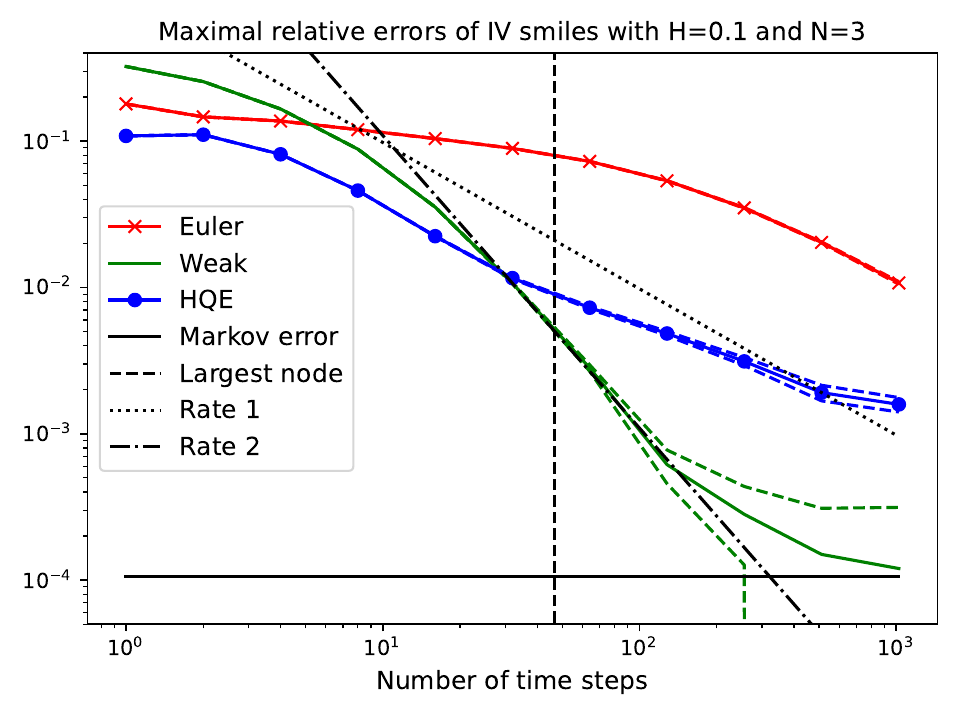}
\end{minipage}
\caption{Errors of implied volatility smiles with $H=0.1$, and $N=2$ (left) or $N=3$ (right). The horizontal black line is the error of the Markovian approximation, while the vertical black line is the largest node $x$ of the Markovian approximation. The solid (blue/red/green) lines represent the maximal errors between the true smile and the smile using simulation. The dashed lines indicate the $95\%$ MC confidence intervals of these errors.}
\label{fig:ErrorsSmileHIs01}
\end{figure}

The results for $H=0.1$ are reported in Table \ref{tab:ErrorsSmileHIs01}, and illustrated in Figure \ref{fig:ErrorsSmileHIs01}. Based on these computations, it seems likely that the HQE scheme and the Euler scheme converge with rate 1 as $M\to\infty$, while the weak scheme converges with rate 2. Furthermore, by comparing $N=2$ with $N=3$, we see that using a higher dimension leads (initially) to slower convergence. This is likely caused by the larger nodes (i.e. higher mean-reversions) that are present in higher dimensions $N$. Indeed, the weak scheme seems to need about $M\ge 16$ (for $N=2$) or $M\ge 64$ (for $N=3$) time steps to get close to its expected convergence rate of 2, which is roughly the size of the largest nodes of $8.7171$ and $46.831$, respectively. The Euler scheme might need a higher $M$ to get close to achieving rate 1, but the general picture is similar. Hence, when choosing the dimension $N$ there is a clear trade-off between a more accurate Markovian approximation, and the less accurate simulation. Finally, to achieve a total relative error of about $1\%$, we need roughly $M=40$ for the HQE scheme (computed by log-linear interpolation of the errors in Table \ref{tab:ErrorsSmileHIs01}), compared to $M=282$ for the Euler scheme (with $N=2$), and $M=13$ for the Weak scheme.

\begin{table}[!htbp]
\centering
\resizebox{\textwidth}{!}{\begin{tabular}{c|c|c|c|c|c|c|c|c|c|c}
     &   \multicolumn{2}{c|}{}  &               \multicolumn{4}{c|}{$N=2$}               & \multicolumn{4}{c}{$N=3$}\\ \hline
     & \multicolumn{2}{c|}{HQE} & \multicolumn{2}{c|}{Euler} & \multicolumn{2}{c|}{Weak} & \multicolumn{2}{c|}{Euler} & \multicolumn{2}{c}{Weak}\\ \hline
$M$  & Error &  Time  & Error &  Time  & Error &  Time  & Error &      Time          & Error & Time\\ \hline
1    & 10.85 & 140.02 & 17.95 & 13.361 & 25.10 & 31.410 & 17.95 & 16.945 & 32.36 & 31.761\\
2    & 11.06 & 169.16 & 14.86 & 22.093 & 15.94 & 51.663 & 14.64 & 24.979 & 25.50 & 58.088\\
4    & 8.124 & 241.56 & 13.92 & 39.432 & 7.161 & 104.74 & 13.71 & 43.509 & 16.58 & 109.58\\
8    & 4.591 & 418.66 & 12.14 & 74.123 & 2.397 & 201.80 & 11.98 & 81.089 & 8.813 & 213.27\\
16   & 2.240 & 803.96 & 9.671 & 149.11 & 0.666 & 384.97 & 10.40 & 154.49 & 3.544 & 424.29\\
32   & 1.158 & 1680.4 & 6.774 & 294.06 & 0.178 & 790.95 & 8.919 & 341.28 & 1.068 & 833.02\\
64   & 0.727 & 3771.5 & 4.162 & 572.43 & 0.053 & 1551.1 & 7.265 & 641.49 & 0.282 & 1655.0\\
128  & 0.485 & 9101.3 & 2.255 & 1147.3 & 0.020 & 3280.9 & 5.358 & 1287.2 & 0.062 & 3685.8\\
256  & 0.312 & 28538  & 1.113 & 2358.0 & 0.020 & 6283.3 & 3.495 & 2831.2 & 0.028 & 7173.7\\
512  & 0.191 & 85816  & 0.514 & 4825.8 & 0.016 & 13172  & 2.031 & 5413.5 & 0.016 & 14370\\
1024 & 0.159 & 331835 & 0.222 & 9686.1 & 0.013 & 26459  & 1.047 & 11044  & 0.012 & 28867
\end{tabular}}
\caption{Maximal (over the strike) relative errors in $\%$ for the implied volatility smiles, where $H=0.1$, together with the computational times in seconds. The error of the Markovian approximation is $0.0131\%$ for $N=2$ and $0.0105\%$ for $N=3$. The MC errors are all at most $0.02\%$, where we used $m=25\cdot 2^{22}$ samples. We remark that we used RQMC for the errors, but MC for the computational times. This is because (R)QMC required us to first simulate and store all necessary random variables, while in MC, we can just generate them once they are needed. Hence, RQMC requires a lot more memory, and as a consequence the computational times for RQMC increased more than linearly (resp. quadratically) for large $M$.}
\label{tab:ErrorsSmileHIs01}
\end{table}

\begin{figure}
\centering
\begin{minipage}{.5\textwidth}
  \centering
  \includegraphics[width=\linewidth]{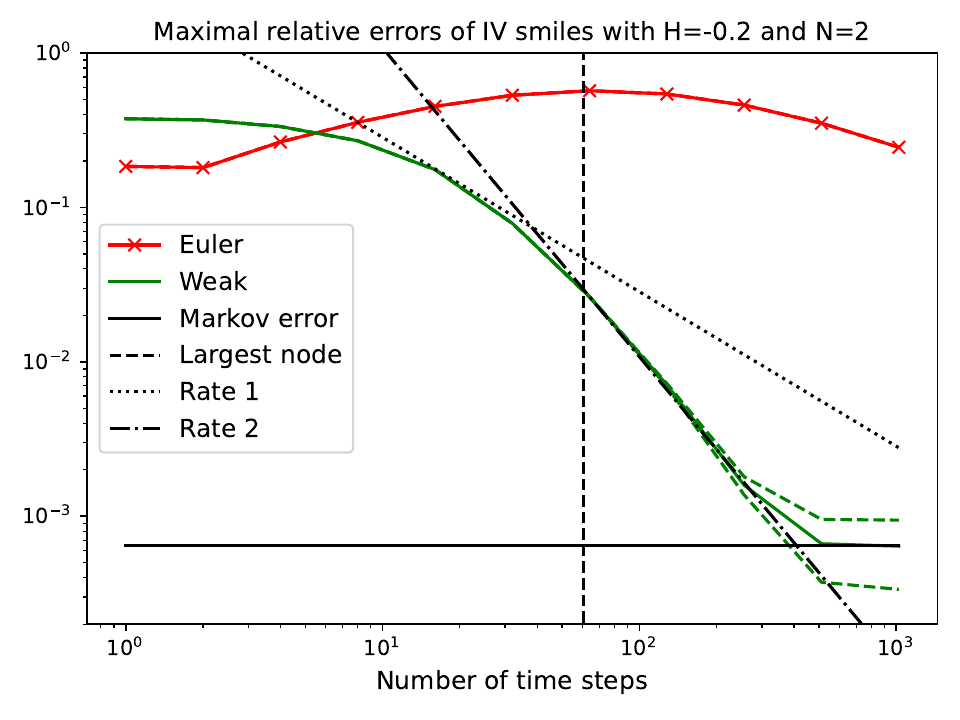}
\end{minipage}%
\begin{minipage}{.5\textwidth}
  \centering
  \includegraphics[width=\linewidth]{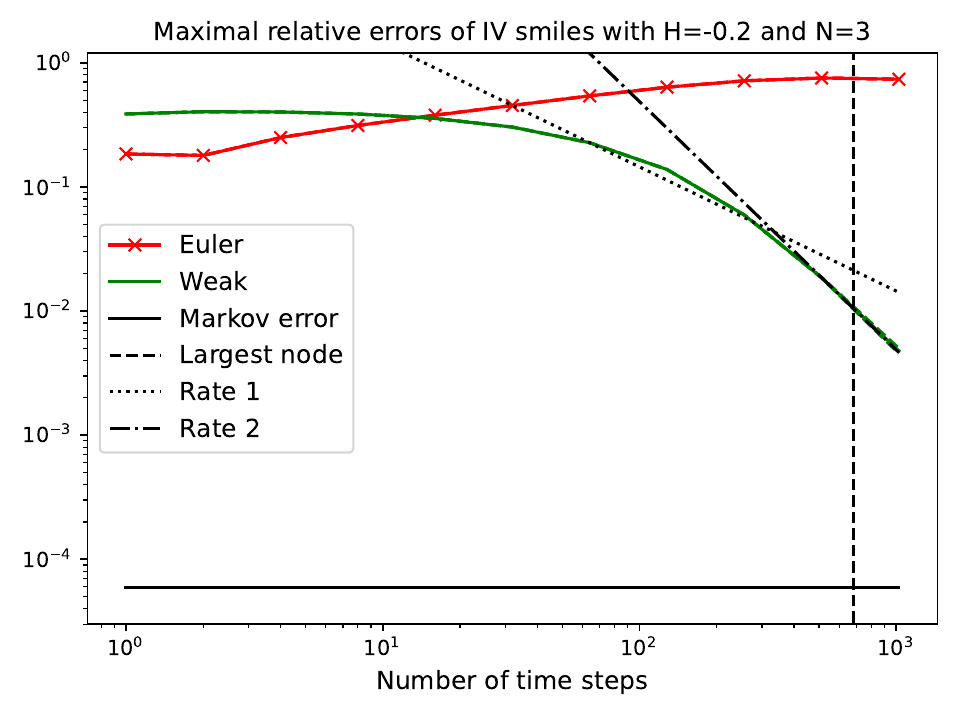}
\end{minipage}
\caption{Errors of implied volatility smiles with $H=-0.2$, and $N=2$ (left) or $N=3$ (right). The horizontal black line is the error of the Markovian approximation, while the vertical black line is the largest node $x$. The solid (red/green) lines represent the maximal errors between the true smile and the smile using simulation. The dashed lines indicate the $95\%$ MC confidence intervals of these errors.}
\label{fig:ErrorsSmileHIsMinus02}
\end{figure}

The results for the Hurst parameter $H=-0.2$ are given in Table \ref{tab:ErrorsSmileHIsMinus02} and Figure \ref{fig:ErrorsSmileHIsMinus02}. This is a much more challenging regime, where the rough Heston model is not well-defined anymore in a strong sense, but merely in a weak sense, see e.g. \cite{abi2023reconciling, jusselin2020no}. Conversely, the Markovian approximations are still standard SDEs. Nonetheless, as was also shown in \cite{bayer2023++Weak}, a higher number of nodes $N$ is necessary to get a similarly small Markovian approximation error as for larger $H$. Additionally, even if we choose the same dimension $N$ as for larger $H$, we still get larger nodes, since larger mean-reversions correspond to the more pronounced singularity in the kernel $K$. Both of these problems combined make the task of simulation much harder, as we correspondingly need more time steps $M$ to achieve small errors. Aside from that, the general picture is similar to the Hurst parameter $H=0.1$, in that the weak scheme (eventually) converges with rate 2, while the Euler scheme likely will converge with rate 1. The weak scheme needs about $M\ge 128$ (for $N=2$) and $M\ge 1024$ (for $N=3$) time steps to get close to rate $2$, which is again similar to the largest nodes of $60.452$ and $681.37$, respectively. To achieve a relative error of about $1\%$, we need about $M=107$ time steps for the weak scheme with $N=2$, while we need $M \gg 512$ for the Euler scheme.

\begin{table}[!htbp]
\centering
\begin{tabular}{c|c|c|c|c}
     & \multicolumn{2}{c|}{$N=2$} & \multicolumn{2}{c}{$N=3$}\\ \hline
$M$  & Euler & Weak  & Euler & Weak\\ \hline
1    & 18.44 & 37.52 & 18.44 & 38.74\\
2    & 18.10 & 36.86 & 17.94 & 40.39\\
4    & 26.52 & 33.41 & 25.00 & 40.19\\
8    & 35.55 & 27.02 & 31.27 & 38.70\\
16   & 45.07 & 17.58 & 37.87 & 35.64\\
32   & 53.27 & 7.885 & 45.33 & 30.36\\
64   & 56.99 & 2.636 & 54.13 & 22.71\\
128  & 54.27 & 0.708 & 63.63 & 13.83\\
256  & 45.98 & 0.159 & 71.63 & 5.945\\
512  & 35.11 & 0.066 & 75.51 & 1.859\\
1024 & 24.50 & 0.064 & 73.81 & 0.478
\end{tabular}
\caption{Maximal (over the strike) relative errors in $\%$ for the implied volatility smiles, where $H=-0.2$. The error of the Markovian approximation is $0.0649\%$ for $N=2$ and $0.00593\%$ for $N=3$. The MC errors are all at most $0.02\%$, where we used $m=25\cdot 2^{22}$ samples.}
\label{tab:ErrorsSmileHIsMinus02}
\end{table}

\subsubsection{Implied volatility surface}\label{sec:EuropeanCallSurfaces}

We now consider an entire volatility surface. Under our standard parameters, we consider the maturities and log-strikes $$T^{(i)} = \frac{i}{16},\qquad \bm{k}^{(i)} = \left\{-0.10, -0.09, \dots, 0.05\right\} \cdot \sqrt{T^{(i)}},\qquad i = 1,\dots,16.$$ Furthermore, we use Markovian approximations with $N=2$ and $N=3$ dimensions. The associated nodes and weights using the BL2 quadrature rule for the above maturity vector are given in Table \ref{tab:QuadratureRuleSurface}. We remark that since we are given an entire vector of maturities, we adapted the BL2 algorithm as recommended in \cite[Section 3.3.2]{bayer2023++Weak}.

\begin{figure}
\centering
\begin{minipage}{.5\textwidth}
  \centering
  \includegraphics[width=\linewidth]{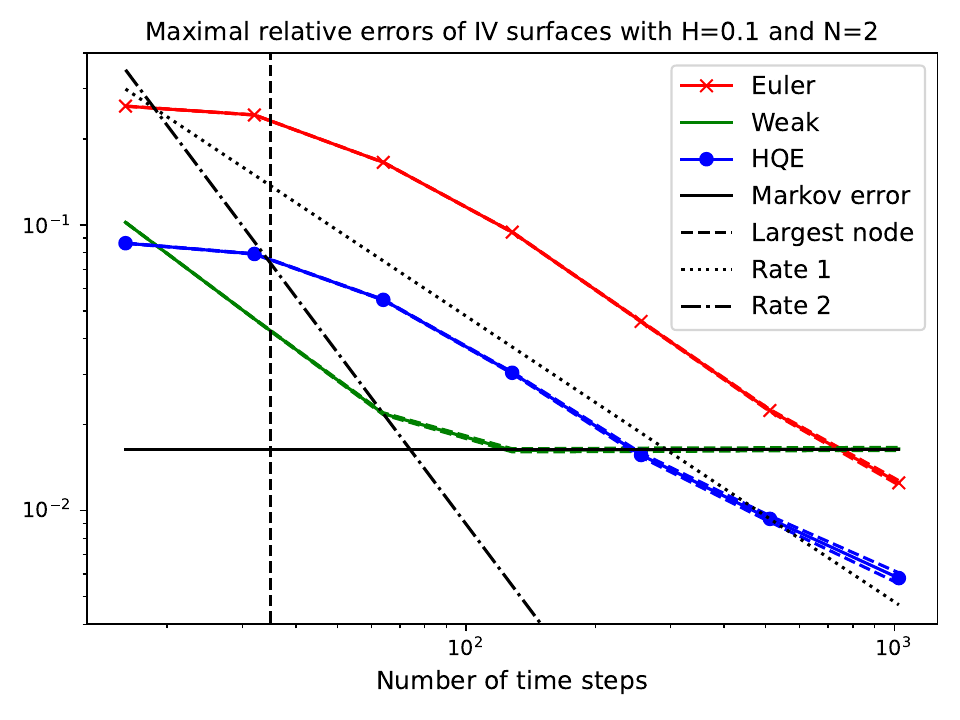}
\end{minipage}%
\begin{minipage}{.5\textwidth}
  \centering
  \includegraphics[width=\linewidth]{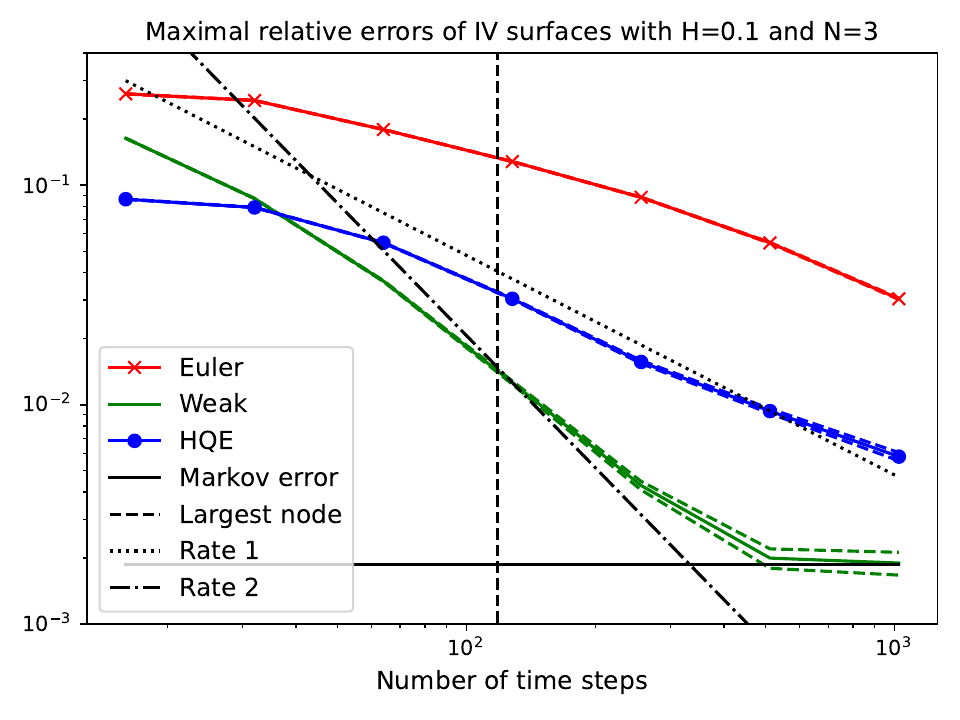}
\end{minipage}
\caption{Errors of implied volatility surfaces with $H=0.1$ for $N=2$ (left) or $N=3$ (right) dimensions. The horizontal black line is the error of the Markovian approximation, while the vertical black line is the largest node $x$. The solid (blue/red/green) lines represent the maximal errors between the true surface and the surface using simulation. The dashed lines indicate the $95\%$ MC confidence intervals of these errors.}
\label{fig:ErrorsSurface}
\end{figure}

The results are given in Table \ref{tab:ErrorsSurface} and Figure \ref{fig:ErrorsSurface}. These algorithms generally show similar results as for the implied volatility smiles. The HQE and the Euler scheme seem to converge roughly with rate $1$, while the Weak scheme approaches rate $2$ for $M\ge 128$ and $M\ge 256$ time steps for $N=2$ and $N=3$, respectively. We compare this with the largest mean-reversions of $35$ and $118$, respectively. Using logarithmic interpolation, the HQE scheme achieves an error of $1\%$ for $M=468$ time steps, the Weak scheme with $N=3$ for $M=149$, and the Euler scheme needs $M=6035$ (by extrapolation).

\begin{table}
\centering
\begin{minipage}{.5\textwidth}
  \centering
  \begin{tabular}{c|c|c}
Nodes & 0.20000 & 34.868\\ \hline
Weights & 1.3360 & 5.6228
\end{tabular}
\end{minipage}%
\begin{minipage}{.5\textwidth}
  \centering
  \begin{tabular}{c|c|c|c}
Nodes & 0.083995 & 5.6485 & 118.01\\ \hline
Weights & 0.80386 & 1.6079 & 8.8078
\end{tabular}
\end{minipage}
\caption{Nodes and weights from the algorithm ``BL2'' in \cite{bayer2023++Weak} for $H=0.1$, the vector of maturities $T = (1/16, 2/16,\dots, 1)$, and $N=2,3$.}
\label{tab:QuadratureRuleSurface}
\end{table}

\begin{table}[!htbp]
\centering
\begin{tabular}{c|c|c|c|c|c}
     &       & \multicolumn{2}{c|}{$N=2$} & \multicolumn{2}{c}{$N=3$}\\ \hline
$M$  &  HQE  & Euler & Weak  & Euler & Weak\\ \hline
16   & 8.631 & 26.06 & 10.21 & 26.06 & 16.35\\
32   & 7.914 & 24.27 & 4.686 & 24.31 & 8.689\\
64   & 5.467 & 16.57 & 2.183 & 17.92 & 3.656\\
128  & 3.036 & 9.440 & 1.625 & 12.81 & 1.261\\
256  & 1.565 & 4.593 & 1.632 & 8.801 & 0.428\\
512  & 0.935 & 2.240 & 1.639 & 5.462 & 0.200\\
1024 & 0.581 & 1.250 & 1.643 & 3.036 & 0.190
\end{tabular}
\caption{Maximal (over the strikes and maturities) relative errors in $\%$ for the implied volatility surfaces. The error of the Markovian approximation is $1.630\%$ for $N=2$ and $0.1867\%$ for $N=3$. The MC errors are all at most $0.02\%$, where we used $m=25\cdot 2^{22}$ samples.}
\label{tab:ErrorsSurface}
\end{table}

\subsubsection{Geometric Asian call option}\label{sec:GeometricAsian}

As a first path-dependent option, we  now consider geometric Asian call options. The advantage of geometric Asian options in the rough Heston model is that the Fourier transform of the random variable $\int_0^T \log(S_t) \sdd t$ is again known in semi-explicit form, so we can again efficiently compute reference prices and compare them to our simulation results. Under our standard parameters, we consider the maturity $T=1$ and log-strikes $$\bm{k} = \left\{-0.10, -0.09, \dots, 0.05\right\}.$$ Again, we use Markovian approximations with $N=2, 3$ dimensions. Since we have the single maturity $T=1$, we use the same nodes and weights as in Table \ref{tab:QuadratureRuleSmile} for the implied volatility smiles.

\begin{figure}
\centering
\begin{minipage}{.5\textwidth}
  \centering
  \includegraphics[width=\linewidth]{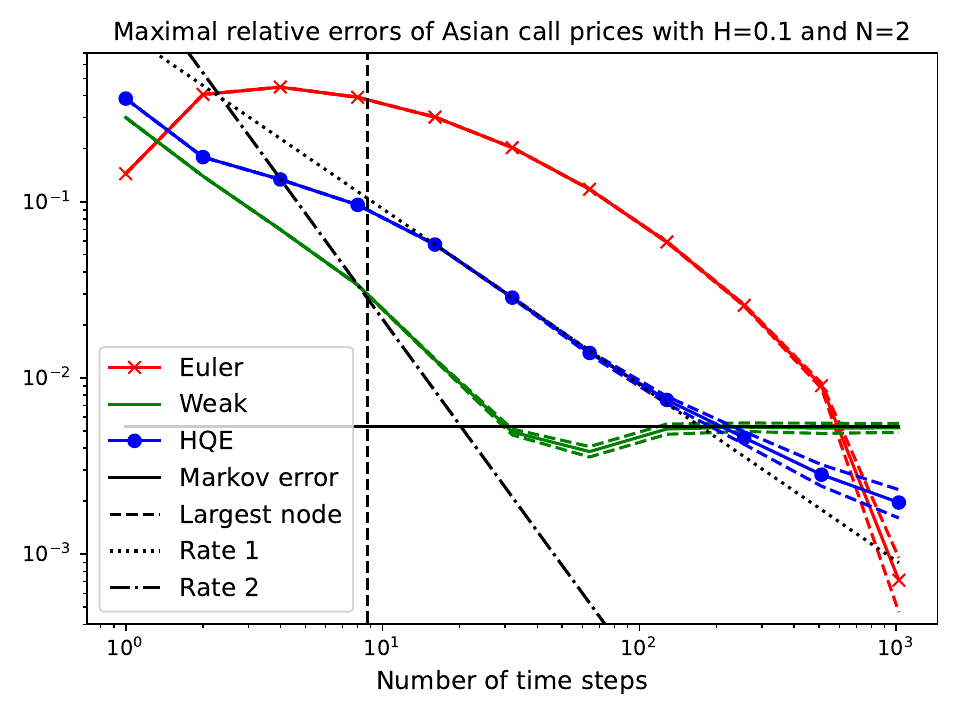}
\end{minipage}%
\begin{minipage}{.5\textwidth}
  \centering
  \includegraphics[width=\linewidth]{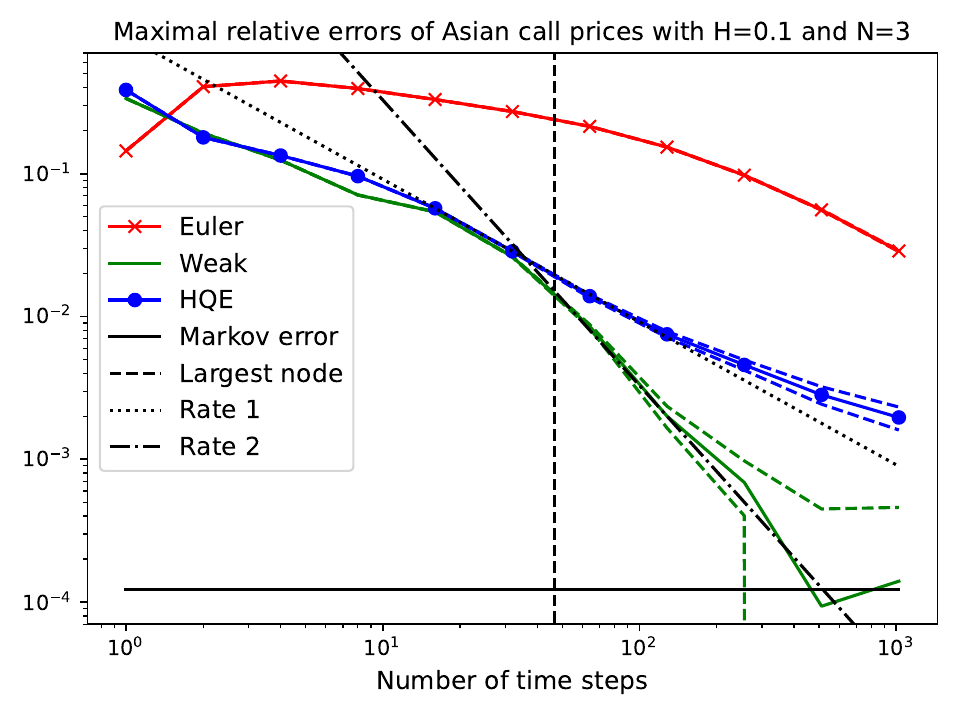}
\end{minipage}
\caption{Errors of prices for Asian call options with $H=0.1$ for $N=2$ (left) or $N=3$ (right) dimensions. The horizontal black line is the error of the Markovian approximation, while the vertical black line is the largest node $x$. The solid (blue/red/green) lines represent the maximal errors between the true prices and the prices using simulation. The dashed lines indicate the $95\%$ MC confidence intervals of these errors.}
\label{fig:ErrorsAsian}
\end{figure}

Given the sample paths we compute the Asian option prices using the trapezoidal rule. The results are given in Table \ref{tab:ErrorsAsian} and Figure \ref{fig:ErrorsAsian}. Note that the weak scheme does not outperform the HQE scheme as much as in the previous examples. This may be because we compute the average log-stock prices using the trapezoidal rule, which introduces an additional discretization error. Nonetheless, the weak scheme is of course still much faster than the HQE scheme given the same number of time steps $M$.

Again, the HQE scheme and the Euler scheme seem to converge with rate 1, where as the weak scheme approaches rate 2 for $M\ge 64$ and $M\ge 128$ time steps for $N=2$ and $N=3$, respectively.  This is a bit larger than the largest nodes of $12.328$ and $59.003$, respectively, possibly also due to the discretization error of the trapezoidal rule. Using logarithmic interpolation, the HQE scheme needs about $M=91$ time steps to achieve a relative error of $1\%$, while the weak scheme needs $M=25$ and the Euler scheme needs $M=789$ (by extrapolation) for $N=2$.

\begin{table}[!htbp]
\centering
\begin{tabular}{c|c|c|c|c|c}
     &       & \multicolumn{2}{c|}{$N=2$} & \multicolumn{2}{c}{$N=3$}\\ \hline
$M$  &  HQE  & Euler & Weak  & Euler & Weak\\ \hline
1    & 38.58 & 14.44 & 30.12 & 14.44 & 33.64\\
2    & 17.95 & 40.72 & 13.98 & 40.66 & 19.22\\
4    & 13.40 & 44.82 & 6.958 & 44.57 & 12.44\\
8    & 9.606 & 39.22 & 3.374 & 39.47 & 7.092\\
16   & 5.722 & 30.29 & 1.270 & 33.05 & 5.402\\
32   & 2.858 & 20.32 & 0.494 & 27.23 & 2.616\\
64   & 1.387 & 11.79 & 0.382 & 21.43 & 0.853\\
128  & 0.750 & 5.916 & 0.513 & 15.34 & 0.200\\
256  & 0.458 & 2.585 & 0.526 & 9.752 & 0.069\\
512  & 0.283 & 0.903 & 0.519 & 5.576 & 0.009\\
1024 & 0.196 & $0.071^*$ & 0.521 & 2.872 & 0.014
\end{tabular}
\caption{Maximal (over the strikes) relative errors in $\%$ for the Asian call prices. The error of the Markovian approximation is $0.5292\%$ for $N=2$ and $0.0123\%$ for $N=3$. The MC errors are all at most $0.03\%$, where we used $m=25\cdot 2^{22}$ samples. We remark that the small error ${}^*$ is the result of a cancellation between the Markovian error of $0.53\%$ and the discretization error of $0.60\%$.}
\label{tab:ErrorsAsian}
\end{table}

\subsubsection{Bermudan put option}\label{sec:BermudanPut}

Next, we value Bermudan put options. For all three schemes (the HQE scheme, the Euler scheme, and the weak scheme), we use the Longstaff-Schwartz algorithm as in \cite{longstaff2001valuing}. In all our examples, we use RQMC with a total of $m=25 \cdot 2^{21}$ samples (i.e. we use 25 random shifts). Moreover, we always use the first $2^{20}$ QMC samples for the linear regression to approximate the stopping rule, and the last $2^{20}$ QMC samples for the pricing of the Bermudan option using this stopping rule.

We discuss the choice of features for the linear regression below. Let us remark right away that the Longstaff-Schwartz algorithm, in its most basic form that we apply, assumes that the underlying process is a Markov process. This is of course given for the Weak scheme and the Euler scheme, since they are simulations of a Markovian approximation of rough Heston. However, this is not given for the HQE scheme, which is a direct simulation of the rough Heston model. Hence, for complete accuracy, one would need to somehow incorporate the history of the sample paths when pricing Bermudan options using the HQE scheme. Since the aim of this paper is to demonstrate the efficacy of simulating the Markovian approximations, and not to develop good methods for pricing Bermudan options using the HQE scheme, we decided to just treat the sample paths simulated by the HQE scheme as Markovian paths, and to not make use of past information.

Throughout these examples, we use the same parameters of the rough Heston model as above, with the exception that we now incorporate a drift $r = 0.06$, and that we use $S_0=100$ for simplicity. Furthermore, throughout we use a Bermudan put option with strike $K=105$. The maturity of the option is $T=1$ (year), and we use two different scenarios, one with 4 exercise times (corresponding to one exercise time per quarter), and one with 16 exercise times (as a proxy for one exercise time per month).

For the Markovian approximations, we use $N=1, 2, 3$ dimensions. The corresponding nodes and weights are the same as in Table \ref{tab:QuadratureRuleSmile} for $N=2,3$. For $N=1$ we have the node $x_1 = 2.1649$ and the weight $w_1 = 2.6233$. We remark that the European put option price under these parameters is roughly $5.244$, and the European put option prices under the Markovian approximations are roughly $5.238$ for $N=1$, and $5.244$ for $N=2, 3$.

\textbf{Choice of features}

One of the important things to consider in the Longstaff-Schwartz algorithm is the choice of features for the linear regression. It seems reasonable to use as features polynomials in the stock price and volatility process. More precisely, before considering polynomials in these processes, we normalize these features by setting $$s \coloneqq \frac{S - K}{K},\quad v\coloneqq V - V_0,\quad \bv = (v^i)_{i=1}^{N-1} = (w_i(V^i - v^i_0))_{i=1}^{N-1}.$$ For the HQE scheme, we of course do not have the features corresponding to $\bv$. Moreover, notice that we drop the $N$-th component $V^N$ in $\bv$. This is because it can already be determined using $V$ and $(V^i)_{i=1}^{N-1}$. Indeed, numerical experiments showed that we gain no additional accuracy if we include the component $V^N$ (but of course the number of features increases, increasing the computational cost).

Using the above variables, it remains to determine which polynomials in $s$, $v$ and $\bv$ we should use. Numerical experiments seem to indicate that the following choice is reasonable: Given a degree $d$, we consider as features all the terms $$s^{d_1} v^{d_2} (v^1)^{d_3}\dots (v^{N-1})^{d_{N+1}}\quad \text{such that}\quad d_1 + 2d_2 + 3d_3 + \dots 3d_{N+1} \le d.$$ In other words, we use polynomials with weighted degree at most $d$, where monomials in $s$ carry the weight 1, monomials in $v$ the weight 2, and monomials in $\bv$ the weight 3. For the HQE scheme we again do not have terms using $\bv$, but we still consider the same weighted degrees for $s$ and $v$. The number of features for various choices of $N$ and $d$ is given in Table \ref{tab:NumberOfFeatures}. The HQE scheme corresponds to $N=1$.

\begin{table}[!htbp]
\centering
\begin{tabular}{c|c|c|c|c|c|c|c|c|c|c}
$d$ & 1 & 2 & 3 & 4 & 5 & 6 & 7 & 8 & 9 & 10\\ \hline
$N=1$ & 1 & 3 & 5 & 8 & 11 & 15 & 19 & 24 & 29 & 35\\
$N=2$ & 1 & 3 & 6 & 10 & 15 & 22 & 30 & 40 & 52 & 66\\
$N=3$ & 1 & 3 & 7 & 12 & 19 & 30 & 43 & 60 & 83 & 110
\end{tabular}
\caption{Number of features for the Longstaff-Schwartz algorithm for various dimensions $N$ (horizontal) of the Markovian approximation and maximal degrees $d$ (vertical) of the feature polynomials. The HQE scheme corresponds to $N=1$.}
\label{tab:NumberOfFeatures}
\end{table}

Finally, it remains to choose an appropriate maximal degree $d$. To determine $d$, we computed the Bermudan prices for the HQE, the Euler, and the Weak scheme using $n=256$ simulation time steps, and $N=3$ dimensions. The resulting prices are given in Table \ref{tab:BermudanPricesDependingOnD}. This table seems to indicate that $d=6$ is a reasonable choice. Indeed, for larger $d$, the results do not markedly improve (i.e. increase) anymore, and overfitting may even occur. Hence, moving forward, we fix $d=6$.

\begin{table}[!htbp]
\centering
\begin{tabular}{c|c|c|c|c|c|c}
 & \multicolumn{3}{c|}{4 execution times} & \multicolumn{3}{c}{16 execution times}\\ \hline
$d$ & HQE & Euler & Weak & HQE & Euler & Weak\\ \hline
1 & 5.881 & 6.055 & 5.887 & 5.976 & 6.148 & 5.983\\
2 & 5.969 & 6.160 & 5.989 & 6.112 & 6.301 & 6.144\\
3 & 5.999 & 6.202 & 6.039 & 6.134 & 6.360 & 6.207\\
4 & 6.006 & 6.217 & 6.055 & 6.151 & 6.388 & 6.231\\
5 & 6.006 & 6.219 & 6.056 & 6.155 & 6.394 & 6.237\\
6 & 6.006 & 6.219 & 6.053 & 6.158 & 6.397 & 6.240\\
7 & 6.008 & 6.219 & 6.055 & 6.159 & 6.397 & 6.244\\
8 & 6.009 & 6.221 & 6.057 & 6.159 & 6.399 & 6.248\\
9 & 6.009 & 6.218 & 6.055 & 6.162 & 6.399 & 6.243\\
10 & 6.009 & 6.217 & 6.054 & 6.159 & 6.397 & 6.239
\end{tabular}
\caption{Prices of the Bermudan put option for various maximal degrees $d$ of the feature polynomials, where we use $M=256$. The Weak scheme and the Euler scheme use $N=3$.}
\label{tab:BermudanPricesDependingOnD}
\end{table}

\textbf{Convergence results}

\begin{figure}
\centering
\begin{minipage}{.5\textwidth}
  \centering
  \includegraphics[width=\linewidth]{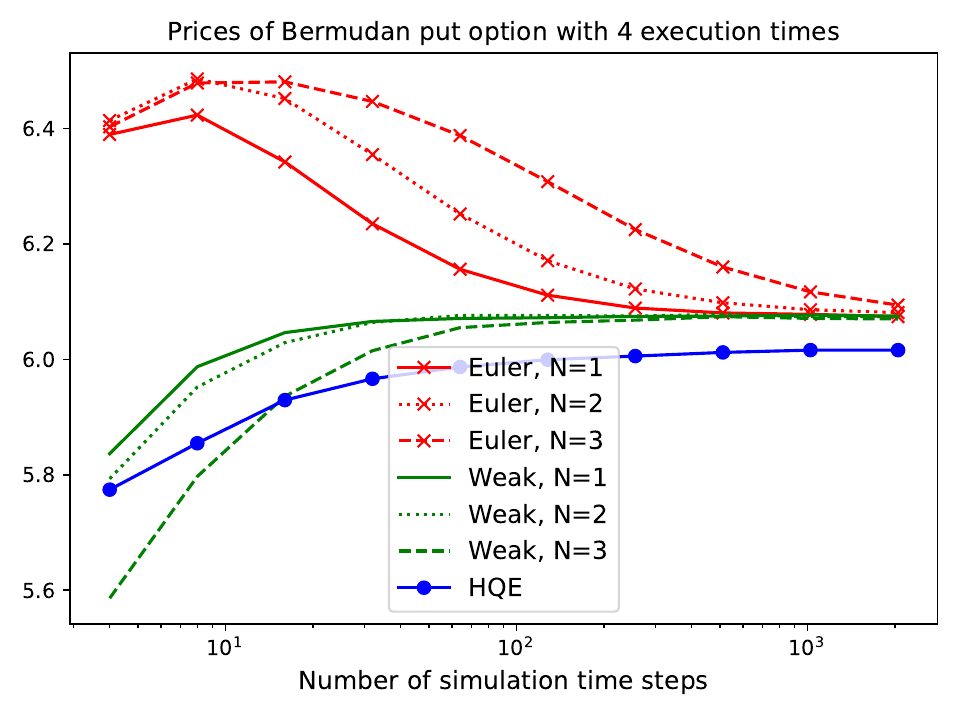}
\end{minipage}%
\begin{minipage}{.5\textwidth}
  \centering
  \includegraphics[width=\linewidth]{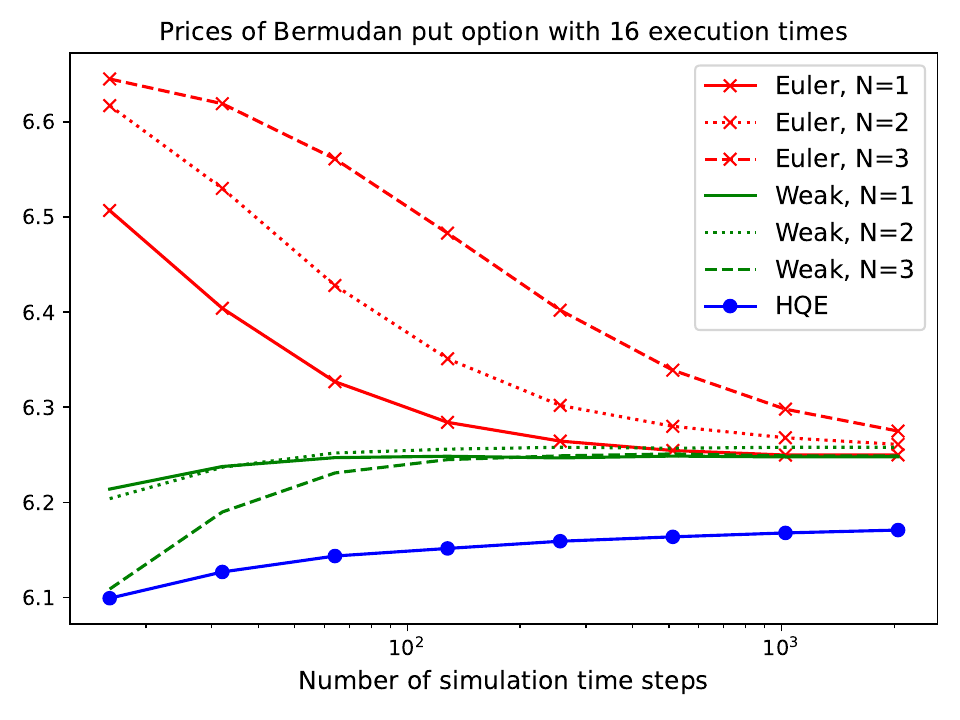}
\end{minipage}
\caption{Bermudan put option prices with $H=0.1$ for $4$ (left) or $16$ (right) exercise times.}
\label{fig:BermudanPrices}
\end{figure}

Having chosen polynomial features with maximal weighted degree $d=6$, we now proceed to pricing the Bermudan put options for various values of approximating dimensions $N$ and number of discretization steps $M$. The results for 4 execution times are in Table \ref{tab:BermudanPrices4}, the results for 16 execution times are in Table \ref{tab:BermudanPrices16}. The option prices are further illustrated in Figure \ref{fig:BermudanPrices}.

\begin{table}[!htbp]
\centering
\begin{tabular}{c|c|c|c|c|c|c|c}
     &  HQE  & \multicolumn{3}{c|}{Euler} & \multicolumn{3}{c}{Weak}\\ \hline
$M$  &       & $N=1$ & $N=2$ & $N=3$ & $N=1$ & $N=2$ & $N=3$\\ \hline
4    & 5.774 & 6.390 & 6.414 & 6.403 & 5.837 & 5.793 & 5.586\\
8    & 5.855 & 6.423 & 6.486 & 6.479 & 5.987 & 5.952 & 5.797\\
16   & 5.929 & 6.342 & 6.452 & 6.481 & 6.046 & 6.029 & 5.936\\
32   & 5.966 & 6.235 & 6.355 & 6.447 & 6.066 & 6.064 & 6.015\\
64   & 5.987 & 6.156 & 6.252 & 6.388 & 6.071 & 6.076 & 6.055\\
128  & 5.999 & 6.111 & 6.171 & 6.308 & 6.072 & 6.076 & 6.064\\
256  & 6.006 & 6.089 & 6.122 & 6.225 & 6.075 & 6.075 & 6.068\\
512  & 6.012 & 6.080 & 6.098 & 6.160 & 6.075 & 6.078 & 6.074\\
1024 & 6.015 & 6.078 & 6.086 & 6.117 & 6.077 & 6.074 & 6.071\\
2048 & 6.016 & 6.074 & 6.081 & 6.094 & 6.075 & 6.074 & 6.070
\end{tabular}
\caption{Prices of the Bermudan put option with 4 execution times depending on the number $M$ of time discretization steps in the simulation. All schemes use feature polynomials up to order $d=6$. The 95$\%$ MC confidence intervals are roughly $0.0025.$}
\label{tab:BermudanPrices4}
\end{table}

\begin{table}[!htbp]
\centering
\begin{tabular}{c|c|c|c|c|c|c|c}
 & HQE & \multicolumn{3}{c|}{Euler} & \multicolumn{3}{c}{Weak}\\ \hline
$M$  &       & $N=1$ & $N=2$ & $N=3$ & $N=1$ & $N=2$ & $N=3$\\ \hline
16   & 6.099 & 6.507 & 6.617 & 6.645 & 6.214 & 6.204 & 6.109\\
32   & 6.127 & 6.404 & 6.530 & 6.619 & 6.238 & 6.237 & 6.190\\
64   & 6.144 & 6.327 & 6.428 & 6.561 & 6.247 & 6.252 & 6.231\\
128  & 6.152 & 6.284 & 6.351 & 6.483 & 6.249 & 6.256 & 6.245\\
256  & 6.159 & 6.264 & 6.302 & 6.402 & 6.247 & 6.258 & 6.249\\
512  & 6.164 & 6.255 & 6.280 & 6.339 & 6.249 & 6.257 & 6.251\\
1024 & 6.168 & 6.250 & 6.268 & 6.298 & 6.248 & 6.258 & 6.249\\
2048 & 6.171 & 6.250 & 6.261 & 6.275 & 6.248 & 6.258 & 6.249
\end{tabular}
\caption{Prices of the Bermudan put option with 16 execution times depending on the number $M$ of time discretization steps in the simulation. All schemes use feature polynomials up to order $d=6$. The 95$\%$ MC confidence intervals are roughly $0.0025.$}
\label{tab:BermudanPrices16}
\end{table}

We can see that both the Weak and the Euler scheme seem to converge to the same value, roughly $6.075$ for 4 execution times and $6.255$ for 16 execution times, both significantly higher than the price of the corresponding European option of $5.244$. In contrast, the HQE scheme converges to the lower values of about $6.02$ and $6.17$, respectively. This difference can likely be explained by the fact that we use the standard Longstaff-Schwartz algorithm for pricing the Bermudan options, and this algorithm assumes the underlying process to be Markov. This is true for the Weak and the Euler scheme, which use the Markovian approximation, but not true for the HQE scheme. Thus, it is natural to expect the HQE scheme to yield lower values, and this also shows that there is relevant information contained in the non-Markovianity of the rough Heston model. This also illustrates a further advantage of the Markovian approximation: Since this approximation yields a Markov process, accurately pricing Bermudan or American options is significantly more straightforward, as one can make use of (usually much simpler) Markovian pricing algorithms.

Finally, let us briefly compare the influence of the approximating dimension $N$ on the computed prices. As before, we can see that the higher $N$ (and thus, the higher the mean-reversions), the longer it takes for the prices to stabilize (as the number of time steps $M$ increases). Second, for 4 execution times there does not seem to be too much of a difference in the computed option prices for large $M$, whereas (at least for the Weak scheme) for 16 execution times $N=2$ seems to yield slightly larger prices than $N=1,3$. On the one hand, $N=2$ more accurately captures the dynamics of the rough Heston model than $N=1$ (in particular the non-Markovianity), possibly explaining the higher values compared to $N=1$. On the other hand, the smaller results for $N=3$ might be the result of overfitting, though this is less clear to the authors and may require further investigation.

\appendix

\printbibliography

\end{document}